\documentclass[12pt]{article}
\usepackage{epsfig,amssymb}

\newcommand{\lsim}{\raisebox{-0.13cm}{~\shortstack{$<$ \\[-0.07cm] $\sim$}}~}
\newcommand{\gsim}{\raisebox{-0.13cm}{~\shortstack{$>$ \\[-0.07cm] $\sim$}}~}

\begin{document}

\begin{flushright}
THES-TP 2001/01 \\
 PM/00-22
\end{flushright}

\vspace{1cm}

\begin{center}

{\large\sc {\bf Optimal Charge and Color Breaking conditions\\
 in the MSSM}}

\vspace{1cm} {C. Le Mou\"el~\footnote{Electronic address:
lemouel@physics.auth.gr}} \vspace{.5cm}

 {Dept. of Theoretical Physics, Aristotle University of
Thessaloniki,\\ GR-54006 Thessaloniki, Greece.}

\vspace{.5cm}

{Physique Math\'ematique et Th\'eorique, UMR No 5825--CNRS, \\
Universit\'e Montpellier II, F--34095 Montpellier Cedex 5,
France.}

\end{center}

\vspace*{1.5cm}

\begin{abstract}
\setlength{\baselineskip}{15pt}

In the MSSM, we make a careful tree-level study of Charge and
Color Breaking conditions in the plane $(H_2, \tilde{u}_L,
\tilde{u}_R)$, focusing on the top quark scalar case. A simple and
fast procedure to compute the VEVs of the dangerous vacuum is
presented and used to derive a model-independent optimal CCB bound
on $A_t$. This bound takes into account all possible deviations of
the CCB vacuum from the D-flat directions. For large $\tan \beta$,
it provides a CCB maximal mixing for the stop scalar fields
$\tilde{t}_1,\tilde{t}_2$, which automatically rules out the Higgs
maximal mixing $|A_t|=\sqrt{6} m_{\tilde{t}}$. As a result, strong
limits on the stop mass spectrum and a reduction, in some cases
substantial, of the one-loop upper bound on the CP-even lightest
Higgs boson mass, $m_h$, are obtained. To incorporate one-loop
leading corrections, this tree-level CCB condition should be
evaluated at an appropriate renormalization scale which proves to
be the SUSY scale.

\end{abstract}

\vspace{.5cm}

\newpage
\setlength{\baselineskip}{15pt}

\section{Introduction}
\setcounter{footnote}{0}

Unlike the Standard Model (SM), the scalar sector of the Minimal
Supersymmetric Standard Model (MSSM) \cite{MSSM} is extremely
large and contains many scalar fields, some of them having
non-trivial color and/or electric charges. The presence of such a
large sector is dictated by supersymmetry (SUSY) \cite{MSSM,SUSY}.
At the Fermi scale, global SUSY must however be broken and soft
SUSY breaking terms which enter mostly in the scalar sector of
MSSM distort the simple analytic structure of the SUSY effective
potential and are responsible of a blowing-up of the number of
free parameters in the MSSM \cite{MSSM,SUSY}. There are many ways
to reduce to a great extent this huge number of parameters,
further improving the predictivity of the MSSM. Each one relies on
some particular scenario of SUSY breaking and mediation to the
MSSM spectrum \cite{SOFT,anom}. Whatever such a model-dependent
scenario may be, phenomenological consistency at the Fermi scale
requires that spontaneous symmetry breaking of the SM gauge group
should occur into the ElectroWeak (EW) vacuum, not in a color
and/or electric charged vacuum. This Charge and Color Breaking
(CCB) danger which does not exist in the SM was quickly realized
in the MSSM \cite{CCB1}, and has been extensively studied ever
since \cite{CCB1,CCB2,CCB3,CCB4,CCB5,CCB6} , providing useful
complementary CCB conditions on the soft parameters.\\
 The major difficulty in obtaining reliable CCB conditions comes from the
extremely involved structure of the effective potential whose
global minimum determines the vacuum of the theory. As a
consequence, CCB studies concentrated on simple directions in the
scalar field space, restricting also often to D-flat directions
\cite{CCB1,CCB2,CCB3,CCB4,CCB5}. The last requirement simplifies
greatly the analytical study of the minima of the potential and
provides already rather strong CCB constraints which may in some
cases rule out model-dependent scenarii \cite{CCB3} or, at least,
severely constrain them \cite{CCB2,CCB3}. Another alternative is
to handle the problem in a purely numerical way \cite{CCB4,CCB6},
a rather blind method, time-consuming, which moreover faces the
danger of missing CCB vacua because of the complexity of the
potential.\\
 Only but a few studies considered analytically, or
semi-analytically, possible deviations of the CCB vacuum from
D-flat directions. In ref.\cite{CCB2}, in particular, it was shown
that in the interesting field planes $(H_1, H_2,\tilde{t}_L,
\tilde{t}_R)$ and $(H_1, H_2,\tilde{t}_L,\tilde{t}_R,
\tilde{\nu}_L)$, such deviations of the CCB vacuum typically
occur, due essentially to large effects induced by the top Yukawa
coupling \cite{CCB2}. To take into account this important feature,
a semi-analytical procedure was proposed, and then illustrated in
an mSUGRA context, giving refined CCB conditions in terms of the
universal soft parameters at the GUT scale \cite{CCB2}. We stress
however that model-dependent assumptions are implicitly present in
this procedure, and need to be relaxed to get a fully satisfactory
model-independent picture of CCB conditions. Furthermore, this
procedure, somewhat, does not lend itself easily to the derivation
of analytical expressions for the CCB Vacuum Expectation Values
(VEVs) of the fields, and, hence, for the optimal conditions to
avoid CCB. On the technical side, our purpose in this paper is to
overcome these difficulties, though in the restricted plane $(H_2,
\tilde{t}_L, \tilde{t}_R)$. This plane will actually provide us
with a simplified framework where to present in detail an
alternative procedure to evaluate the CCB VEVs. This way we will
include in our study all possible deviations from the D-flat
directions, and obtain an accurate analytical information on them.
This procedure can also be adapted to extended planes, and the
present study will be followed by a complete investigation of CCB
conditions in the planes $(H_1, H_2,\tilde{t}_L, \tilde{t}_R)$ and
$(H_1,H_2,\tilde{t}_L, \tilde{t}_R,\tilde{\nu}_L)$
\cite{CCB4champs,CCB5champs}.\\
 More fundamentally, the plane
$(H_2, \tilde{t}_L, \tilde{t}_R)$ is also of particular interest
for the following reasons:\\
 {\bf{i)}} The CCB vacuum typically
deviates largely from the $SU(2)_L \times U(1)_Y$ D-flat
direction, as already observed in \cite{CCB2}, but also from the
$SU(3)_c$ D-flat direction. The latter result, also shared by the
potential in the extended planes
$(H_1,H_2,\tilde{t}_L,\tilde{t}_R)$ and $(H_1,
H_2,\tilde{t}_L,\tilde{t}_R, \tilde{\nu}_L)$
\cite{CCB4champs,CCB5champs}, is in disagreement with the claim of
\cite{CCB2}. As we will see, this important feature must be
incorporated in order to obtain an optimal CCB condition which
encompasses the requirement of avoiding a tachyonic lightest stop.
We will give simple analytic criteria for alignment of the CCB
vacuum in D-flat directions and show that alignment in the
$SU(3)_c$ D-flat direction is in fact a model-dependent statement
which is approximately valid in an mSUGRA scenario \cite{SOFT},
but not in other interesting models, e.g., some string-inspired or
anomaly mediated scenarii \cite{anom}.\\ {\bf{ii)}} For large
$\tan \beta$, the EW vacuum is located in the vicinity of the
plane $(H_2,\tilde{t}_L, \tilde{t}_R)$. Therefore, in this regime,
the study of this plane is self-sufficient: the free parameters
entering the effective potential are enough to evaluate the
optimal necessary and sufficient condition on $A_t$ to avoid CCB.
This does not mean that CCB conditions in the plane
$(H_2,\tilde{t}_L, \tilde{t}_R)$ are useless for low $\tan \beta$.
We will give in this paper an analytic optimal sufficient
condition to avoid CCB in this plane, and, to evaluate the
complementary optimal necessary CCB condition, we will simply need
some additional information on the depth of the EW potential,
which reduces in fact to a particular choice for $\tan \beta$ and
the pseudo-scalar mass $m_{A^0}$.\\ Our purpose in this study is
also to consider some physical consequences at the SUSY scale of
the CCB conditions. We will investigate in detail the benchmark
scenario $M_{SUSY}=m_{\tilde{t}_L}=m_{\tilde{t}_R}$ and $\tan
\beta = + \infty$, often considered in Higgs phenomenology
\cite{higgs,higgsrev}. In this case, the stop mixing parameter
equals the trilinear soft term, $\tilde{A}_t \equiv A_t+\mu / \tan
\beta =A_t$. We will show that the so-called Higgs maximal mixing
$|A_t|=\sqrt{6} m_{\tilde{t}}$ is always largely ruled out by the
optimal CCB condition. This will lead us to introduce a CCB
maximal mixing, which induces strong bounds on the stop mass
spectrum. Another direct implication of this result is a lowering
of the one-loop upper bound on the CP-even lightest Higgs boson
mass $m_h$ reached for such a large $\tan \beta$ regime
\cite{higgs, higgsrev}. For illustration, this point will be
considered in a simplified setting, where only top and stop
contributions to $m_h$ will be taken into account, assuming
$m_{A^0} \gg m_{Z^0}$. This will already point out the importance
of CCB conditions in this context, but should however be
completed, to become more realistic, by a refined investigation
including all one-loop and two loop contributions to $m_h$
\cite{higgs,higgsrev}. In these illustrations, the leading
one-loop corrections to the CCB condition will be incorporated by
assuming that the tree-level CCB condition obtained are evaluated
at an appropriate renormalization scale, estimated in fact to be
the SUSY scale \cite{CCB2,V1Q}. This way, we expect the results
presented in this paper to be robust under inclusion of such
radiative corrections. Finally, we note that these results can be
shown to be also numerically representative of the large $\tan
\beta$ regime with small enough values of the supersymmetric term
$\mu$, i.e. $\tan \beta \gsim 15$ and $|\mu| \lsim
Min[m_{A^0},M_{SUSY}]$ \cite{CCB4champs}.\\ {\bf{iii)}} As is
well-known, for metastability considerations, CCB vacua associated
with the third generation of squarks are the most dangerous ones
\cite{CCB5,CCB6,meta}. This comes from the fact that such vacua
prove to be rather close to the EW vacuum, resulting in a barrier
separating both vacua more transparent to a tunneling effect. We
will see that combining experimental data on the lower bound of
the lightest stop mass, $m_{\tilde{t}_1}$, with precise CCB
conditions already delineates large regions in the parameter space
where the EW vacuum is the deepest one and, hence, stable. Outside
such regions, an optimal determination of the modified CCB
metastable conditions requires first a precise knowledge of the
geometrical properties of the effective potential, e.g., the
positions of the CCB vacua and saddle-points. In this light, the
analytical expressions presented in this article provide an
essential information to investigate precisely metastability.\\

The paper is organized as follows. In section 2, we first review
the issue of CCB conditions in the plane $(H_2,\tilde{u}_L,
\tilde{u}_R)$ in the D-flat direction. We turn then to the full
plane case for the third generation of squark fields and give a
first simple analytical sufficient condition on $A_t$ to avoid
CCB. In section 3, we detail our semi-analytical procedure to
obtain the VEVs of the local extrema in this plane, discuss the
deviation from the $SU(3)_c$ D-flat direction, and give an optimal
sufficient bound on $A_t$ to avoid CCB. We discuss finally some
geometrical features of the CCB vacuum. In section 4, we discuss
the renormalization scale at which the tree-level CCB conditions
obtained should be evaluated in order to incorporate leading
one-loop corrections. In section 5, we summarize the practical
steps needed to evaluate numerically the optimal necessary and
sufficient CCB condition on $A_t$. Sections 6-7 are devoted to
numerical illustrations and, for large $\tan \beta$, to
phenomenological implications of the new optimal CCB condition for
the stop mass spectrum and the one-loop upper bound on the
lightest Higgs boson  mass. Section 8 presents our conclusions.
Finally, the appendices A and B contain some technical material
and the generalization of this study to the plane $(H_1,
\tilde{b}_L, \tilde{b}_R)$, valid for a large bottom Yukawa
coupling, or equivalently for large $\tan \beta$.

\section{CCB conditions in the plane $(H_2, \tilde{u}_L, \tilde{u}_R)$}

We consider the tree-level effective potential in the plane
$(H_2,\tilde{u}_L, \tilde{u}_R)$, where $H_2$ denotes the neutral
component of the corresponding Higgs scalar $SU(2)_L$ doublet, and
$\tilde{u}_L$, $\tilde{u}_R$ are respectively the left and right
up squark of the same generation. In this plane, the tree-level
effective potential reads \cite{CCB1}
\begin{eqnarray}
\label{V3ch} V_3&=&m_2^2 H_2^2+m_{\tilde{u}_L}^2 {\tilde{u}_L}^2+
m_{\tilde{u}_R}^2 {\tilde{u}_R}^2
 -2 Y_{u} A_u H_2 \tilde{u}_L \tilde{u}_R
+Y_u^2 (H_2^2 {\tilde{u}_L}^2+H_2^2 {\tilde{u}_R}^2+
{\tilde{u}_L}^2 {\tilde{u}_R}^2) \nonumber \\ &&
+{\frac{g_1^2}{8}} (H_2^2+{\frac{{\tilde{u}_L}^2}{3}}- {\frac{4
{\tilde{u}_R}^2}{3}})^2+{\frac{g_2^2}{8}}
(H_2^2-{\tilde{u}_L}^2)^2+ {\frac{g_3^2}{6}}
({\tilde{u}_L}^2-{\tilde{u}_R}^2)^2
\end{eqnarray}
We suppose that all fields are real and that $H_2,\tilde{u}_L$ are
positive, which can be arranged by a phase redefinition. The Higgs
mass parameter $m_2^2=m_{H_2}^2+\mu^2$ can have both signs,
$m_{H_2}$ being the soft mass of the corresponding Higgs field;
$m_{\tilde{u}_L}^2$,$ m_{\tilde{u}_R}^2$ are the squared soft
masses of the left and right up squarks and are supposed to be
positive to avoid instability of the potential at the origin of
the fields; $Y_u$ and $A_u$ stand for the Yukawa coupling and the
trilinear soft coupling and are also supposed to be real and
positive, which can be arranged once again by a phase redefinition
of the fields; finally $g_1,g_2,g_3$ are respectively the $U(1)_Y,
SU(2)_L, SU(3)_c$ gauge couplings.

\subsection{The D-flat direction}

In the D-flat direction $|H_2|=|\tilde{u}_L|=|\tilde{u}_R|$, the
potential $V_3$, eq.(\ref{V3ch}), may develop a very deep CCB
minimum, unless the well-known condition \cite{CCB1,CCB2}
\begin{equation}
\label{condfrere} A_u^2 \le (A_{u,3}^D)^2 \equiv 3
(m_{\tilde{u}_L}^2+m_{\tilde{u}_R}^2+m_2^2)
\end{equation}
is verified. Strictly speaking, as the extremal equations easily
show, the global minimum of the potential $V_3$, eq.(\ref{V3ch}),
lies in the D-flat direction only for:
\begin{equation}
\label{massrel} m_{\tilde{u}_L}^2=m_{\tilde{u}_R}^2=m_2^2
\end{equation}
However, {\sl in the small Yukawa coupling regime}, valid for the
first two generations of quarks, the VEVs of the CCB vacuum are
large, $< \phi
> \sim A_u / 3 Y_u$. The vacuum then proves to be located in the
vicinity of this direction \cite{CCB1,CCB2}, even for large
deviations from the mass relations in eq.(\ref{massrel}).
Moreover, due to the smallness of the Yukawa coupling, this CCB
minimum is very deep, $<V_3> \lsim -A_u^2 [A_u^2-(A_{u,3}^D)^2]/27
Y_u^2$, and, with increasing $A_u$, gets rapidly\footnote{ Within
less than 1 GeV.} deeper than the realistic EW vacuum. As a
result, the relation eq.(\ref{condfrere}) turns out to provide an
accurate necessary and sufficient condition to avoid a CCB in this
plane \cite{CCB1,CCB2}.\\

{\sl In the large Yukawa coupling regime}, valid for the top quark
case, the condition eq.(\ref{condfrere}) is now only approximately
necessary, because in some (small) range of values for $A_t$ where
it is not verified the CCB local minimum in the D-flat direction
develops without being deeper than the EW vacuum. {\sl It is
however no more sufficient, the true global CCB minimum of $V_3$,
eq.(\ref{V3ch}), being in general located far away from the D-flat
direction} \cite{CCB2}! Obtaining the most accurate conditions to
avoid CCB in the top quark regime needs to explore the scalar
field space outside D-flat directions, a more difficult task which
is of particular phenomenological interest, as we will see. In the
following, we focus on this interesting regime in order to get a
complete model-independent picture of CCB conditions in the plane
$(H_2,\tilde{t}_L, \tilde{t}_R)$.

\subsection{The full-plane case}

Beyond a critical value for the trilinear soft term $A_t$, a
dangerous CCB minimum, deeper than the EW vacuum, forms and
deepens with increasing values of $A_t$. In ref.\cite{CCB2}, it
was advocated that such a global CCB minimum is located in general
far away from the $SU(2)_L\times U(1)_Y$ D-flat directions, but
close to the $SU(3)_c$ D-flat one. This work was performed in more
extended planes with additional scalar fields, $H_1$ and possibly
a sneutrino field $\tilde{\nu}_L$, which we will consider in
separate articles \cite{CCB4champs, CCB5champs}. Already in the
plane $(H_2,\tilde{t}_L, \tilde{t}_R)$, our study indeed shows a
typical large deviation of the CCB vacuum from the $SU(2)_L\times
U(1)_Y$ D-flat directions. However, in a model-independent way, we
disagree with the claim of ref.\cite{CCB2} that the CCB vacuum
always proves to be located in the vicinity of the $SU(3)_c$
D-flat direction. Actually, alignment in this direction depends on
the magnitude of the soft terms
$A_t,m_{\tilde{t}_L},m_{\tilde{t}_R}$ [see sec.3.2] and occurs in
two different circumstances: either i)
$m_{\tilde{t}_L}=m_{\tilde{t}_R}$, or ii) $A_t \gg
m_{\tilde{t}_L},m_{\tilde{t}_R}$.  Any discrepancy between the
soft masses $m_{\tilde{t}_L},m_{\tilde{t}_R}$, as happens for
instance in some anomaly mediated models \cite{anom}, is the
source of a possibly large departure of the CCB vacuum from the
$SU(3)_c$ D-flat direction and, ultimately, of a sizeable
enhancement of the optimal condition on $A_t$ to avoid CCB.\\
 To consider this feature, we introduce a new parameter which
conveniently measures the separation of the CCB extrema from the
$SU(3)_c$ D-flat direction
\begin{equation}
\label{ft}
 f \equiv {\frac{{\tilde{t}_R}}{{\tilde{t}_L}}}
\end{equation}
Alignment in the $SU(3)_c$ D-flat direction corresponds to
$<f>=\pm1$.\\
 We replace now $\tilde{t}_R \rightarrow f \ \tilde{t}_L$ in
$V_3$, eq.(\ref{V3ch}), which is unambiguous provided $\tilde{t}_L
\neq 0$. By inspection of the extremal equation associated with
the field $\tilde{t}_L$, it is easy to obtain a critical bound on
$A_t$ below which no CCB local
 minimum may exist. The non-trivial solution for the VEV
 $<\tilde{t}_L>$ verifies
\begin{equation}
\label{eqUL} A_3 \ <\tilde{t}_L>^2+ 2 B_3=0
\end{equation}
where $A_3\equiv g_1^2(4 <f>^2-1)^2/18+g_2^2/2 <f>^4+2 g_3^2
(<f>^2-1)^2/3+4 Y_t^2 <f>^2$. This term is a sum of squared terms,
therefore always positive. This implies the inequality
\begin{eqnarray}
\label{B3} B_3 &\equiv &<H_2>^2 \frac{[(12 Y_t^2-4 g_1^2) <f>^2+12
Y_t^2+ g_1^2-3 g_2^2]}{12} \nonumber \\ &&
 -2 A_t Y_t <f> <H_2>+
m_{\tilde{t}_L}^2+<f>^2 m_{\tilde{t}_R}^2 \le 0
\end{eqnarray}
$B_3$ may be considered as a polynomial in $<H_2>$. For $Y_t \ge
Max[\sqrt{(3 g_2^2-g_1^2)/12},g_1/\sqrt{3}]$ the coefficient of
the quadratic term in $<H_2>$ is  positive, whatever $<f>$ is. At
the EW scale, this relation reduces to $Y_t \gsim 0.3$, which
anyway must be verified in order to have a correct top quark mass
$m_t \sim 175 \ GeV$. Obviously, the running of the parameters
$Y_t, g_1,g_2$ with respect to the renormalization scale will not
alter this result, so that we can safely conclude that this
coefficient is always positive. {\sl Actually, this is precisely
the turning point where the qualitative difference between the
large and the small Yukawa coupling regimes enters the game.}\\
 Keeping in mind this
feature, eq.(\ref{B3}) implies that $<f>$ (and thus
$<{\tilde{t}_R}>$) must be positive. The negativity of $B_3$
requires in addition the positivity of the discriminant of $B_3$
considered as a polynomial in $<H_2>$:
\begin{eqnarray}
\label{delB3} \Delta_{B_3}&=& (m_{\tilde{t}_L}^2+<f>^2
m_{\tilde{t}_R}^2) [-4 Y_t^2 (<f>^2+1)+{\frac{g_1^2}{3}} (4
<f>^2-1)+g_2^2] \nonumber \\
 && + 4 A_t^2 Y_t^2 <f>^2 \ge 0
\end{eqnarray}
This inequality may be expressed as a condition on $A_t$
\begin{equation}
A_t^2 \ge -{\frac{(m_{\tilde{t}_L}^2+<f>^2 m_{\tilde{t}_R}^2) [-4
Y_t^2 (<f>^2+1)+{\frac{g_1^2}{3}} (4 <f>^2-1)+g_2^2]}{4 Y_t^2
<f>^2}}
\end{equation}
The right hand side of this relation considered as a function of
$<f>$ is bounded from below and gives an absolute lower bound on
$A_t$ below which $\Delta_{B_3}$ cannot be positive. This lower
bound provides a first very simple sufficient condition to avoid
any CCB minimum in the plane $(H_2, \tilde{t}_L, \tilde{t}_R)$:
\begin{equation}
\label{condsuf} A_t \le A_t^{(0)} \equiv m_{\tilde{t}_L}
\sqrt{1-{\frac{g_1^2}{3 Y_t^2}}}+m_{\tilde{t}_R}
\sqrt{{1-{\frac{(3 g_2^2-g_1^2)}{12 Y_t^2}}}} \sim
m_{\tilde{t}_L}+m_{\tilde{t}_R}
\end{equation}
If this condition is verified, the global minimum of the potential
$V_3$, eq.(\ref{V3ch}), is automatically trapped in the plane
$\tilde{t}_R=\tilde{t}_L=0$ and cannot be lower than the EW
vacuum.\\
 Such a simple relation also sets a lower bound on $A_t$ above
which CCB may possibly occur and is already quite useful to secure
some model-dependent scenarii. As a simple illustration, we
consider the infrared quasi-fixed point scenario for low and large
$\tan \beta$, in an mSUGRA context \cite{IRQFP1,IRQFP2}. For a top
Yukawa coupling large enough at the GUT scale, the soft parameters
$m_{\tilde{t}_L}, m_{\tilde{t}_R}$ and $A_t$ are strongly
attracted in the infrared regime to quasi-fixed points. For
${\frac{Y_t}{g_i}}|_{M_{GUT}}=5$ and $m_0 \lsim m_{1/2}$, where
$g_i$ stands for the three gauge couplings that unify at the GUT
scale, $M_{GUT}$, and $m_0, m_{1/2}$ are respectively the unified
scalar and gaugino masses at the GUT scale, we have at one-loop
level, in the infrared regime \cite{IRQFP2}:
\begin{eqnarray}
&& m_{\tilde{t}_L}^2 \sim 0.70 \ M_3^2 \ \ \ \ \ , \ \ \ \ \
m_{\tilde{t}_R}^2 \sim 0.48 \ M_3^2 \ \ \ \ \  for \ low \ \tan
\beta
\\ && m_{\tilde{t}_L}^2 \sim 0.58 \ M_3^2 \ \ \ \ \ , \ \ \ \ \
 m_{\tilde{t}_R}^2 \sim 0.52 \ M_3^2 \ \ \ \ \  for \ large \ \tan \beta
\end{eqnarray}
 giving $A_t^{(0)} \sim 1.53 \ |M_3|$ ( for low $\tan \beta$) and
$A_t^{(0)} \sim 1.48 \ |M_3|$ ( for large $\tan \beta$), where
$M_3$ is the gluino mass. Comparing these bounds with the infrared
quasi-fixed point value $|A_t| \sim 0.62 \ |M_3|$ ( for both low
and large $\tan \beta$) \cite{IRQFP2}, we see that the sufficient
condition $|A_t| \le A_t^{(0)}$, eq.(\ref{condsuf}), is largely
fulfilled. Therefore, we conclude that the infrared quasi-fixed
point scenario is free of CCB danger in the restricted plane
$(H_2, \tilde{t}_L, \tilde{t}_R)$.

\subsection{The critical CCB condition}

In order to study the CCB extrema of the potential $V_3$,
eq.(\ref{V3ch}), we suppose in the following that $A_t>
A_t^{(0)}$. Consistency requires that, at any CCB extremum, $<f>$
and $<H_2>$ are restricted to intervals which depend essentially
on the soft breaking terms $A_t,m_{\tilde{t}_L},m_{\tilde{t}_R}$:
the positivity of $\Delta_{B_3}$, eq.(\ref{delB3}), which may be
viewed as a polynomial in $<f>^2$, restricts $<f>$ in the interval
given by the real and positive roots of $\Delta_{B_3}$; $<H_2>$
must be included between the real and positive roots of $B_3$,
eq.(\ref{B3}). The potentially dangerous region of positive
$<{\tilde{t}_L}>^2$ then proves to be located in a compact domain
in the plane $(H_2, f)$, growing with $A_t$, with maximal
extension
\begin{eqnarray}
\label{compact1}
 0 &\le & <H_2> \le \frac{6 A_t Y_t} {\sqrt{(12
Y_t^2+g_1^2-3 g_2^2) (3 Y_t^2-g_1^2)}} \lsim \frac{A_t}{Y_t} \\
\label{compact2}
 0 & \le & \ < f> \ \le
{\frac{\sqrt{A_t^2 -m_{\tilde{t}_L}^2 (1-g_1^2/3 Y_t^2)-
m_{\tilde{t}_R}^2(1-(3 g_2^2-g_1^2)/12 Y_t^2)}}{m_{\tilde{t}_R}
\sqrt{1- g_1^2/3Y_t^2}}} \lsim {\frac{A_t}{m_{\tilde{t}_R}}}
\end{eqnarray}
Let us now replace in the potential $V_3$, eq.(\ref{V3ch}),
$\tilde{t}_L$ by the solution of eq.(\ref{eqUL}), and calculate
the remaining two extremal equations. The derivative with respect
to $H_2$ provides an equation cubic in $H_2$ and quartic in $f$
\begin{equation}
\label{eqH23} \alpha_3 H_2^3+ \beta_3 H_2^2+\gamma_3
H_2+\delta_3=0
\end{equation}
with the coefficients
\begin{eqnarray}
\label{alpha}
 \alpha_3 &=&-36 Y_t^4 (f^2+1)^2+[3 g_3^2
(g_1^2+g_2^2)+4 g_1^2 g_2^2] (f^2-1)^2 \nonumber \\ &&+ 6 Y_t^2
g_1^2 (4 f^4+6 f^2-1)+18 Y_t^2 g_2^2 (2 f^2+1)  \\ \label{beta}
 \beta_3&=&9 A_t Y_t f [(12 Y_t^2-4 g_1^2) f^2+12 Y_t^2+g_1^2-3 g_2^2]  \\
 \label{gamma}
  \gamma_3&=&-72 A_t^2 f^2 Y_t^2-3
(m_{\tilde{t}_L}^2+f^2 m_{\tilde{t}_R}^2) [(12 Y_t^2-4 g_1^2)
f^2+12 Y_t^2+g_1^2-3 g_2^2] \nonumber
\\ && + m_2^2 [72 Y_t^2 f^2+g_1^2 (4 f^2-1)^2+9 g_2^2+12 g_3^2
(f^2-1)^2] \\
 \label{delta}
\delta_3 &=&36 A_t Y_t f (m_{\tilde{t}_L}^2+f^2 m_{\tilde{t}_R}^2)
\end{eqnarray}
The derivative with respect to $f$ provides an equation quadratic
in $H_2$ and quartic in $f$
\begin{equation}
\label{eqf3} a_3 f H_2^2+ b_3 H_2+c_3 f=0
\end{equation}
with the coefficients
\begin{eqnarray}
\label{a3}
  a_3&=&2 Y_t^2 [ (18 Y_t^2-12 g_3^2) (f^2-1) +9 g_2^2-
g_1^2 (16 f^2-1)] \nonumber \\
 &&+(f^2-1) [4 g_1^2 g_2^2+3
(g_1^2+g_2^2) g_3^2]  \\
 \label{b3}
 b_3&=& A_t Y_t [12 g_3^2 (f^4-1)-9 g_2^2+
g_1^2 (16 f^4-1)]  \\
 \label{c3}
  c_3&=& - m_{\tilde{t}_L}^2 [ 36
Y_t^2+12 g_3^2 (f^2-1) +4 g_1^2 (4 f^2-1)] \nonumber \\ &&
+m_{\tilde{t}_R}^2 [ 36 f^2 Y_t^2-12 g_3^2 (f^2-1)+9 g_2^2- g_1^2
(4 f^2-1)]
\end{eqnarray}
A dangerous CCB minimum will have to verify the system of coupled
equations eqs.(\ref{eqH23},\ref{eqf3}), with the additional
constraints that $<H_2>,<f>$ should be contained in the compact
domain where $<\tilde{t}_L>^2 \ge 0$ [see eqs.(\ref{compact1},
\ref{compact2})].\\
 To determine if such a CCB
vacuum induces eventually a CCB situation, we need some additional
information on the depth of the potential at a realistic EW
vacuum. In the EW direction, the tree-level potential reads:
\begin{equation}
\label{potEW}
  V|_{EW}= m_1^2 H_1^2+m_2^2 H_2^2-2 m_3^2 H_1
H_2+{\frac{(g_1^2+g_2^2)}{8}} (H_1^2-H_2^2)^2
\end{equation}
where $H_1$ denotes the neutral component of the corresponding
Higgs scalar $SU(2)_L$ doublet and is supposed to be real and
positive, which can be arranged by a phase redefinition of the
fields. Without loss of generality, we may also suppose that the
Higgs mass parameters $m_1^2, m_3^2$ are positive. The extremal
equations in the EW direction read \cite{MSSM}:
\begin{eqnarray}
\label{eqextre}
 && {\frac{m_1^2-m_2^2 \tan^2
\beta}{\tan^2\beta-1}}-{\frac{m_{Z^0}^2}{2}}=0 \\
 \label{eqextre1}
 && (m_1^2+m_2^2) \tan \beta- m_3^2 (1+ \tan^2 \beta)=0
\end{eqnarray}
For $\tan \beta \equiv v_2/v_1 \ge 1$, where $v_1,v_2$ are the
VEVs of the EW vacuum, the minimal value of the EW potential is
given by:
\begin{equation}
\label{VEW}
 <V>|_{EW}=-{\frac{[m_2^2-m_1^2+\sqrt{(m_1^2+m_2^2)^2-4 m_3^4}]^2}{2
(g_1^2+g_2^2)}}
\end{equation}
The realistic EW vacuum is furthermore subject to the
phenomenological constraint $v_1^2+v_2^2=(174 \ GeV)^2$, to
reproduce correct masses for the gauge bosons $Z^0,W^{\pm}$.
Experimental data also completely determine the gauge couplings
$g_1,g_2,g_3$, and the top Yukawa coupling $Y_t$, the latter as a
function of the physical top mass. Besides, we note that the depth
of the EW potential $<V>|_{EW}$, eq.(\ref{VEW}), can be expressed
with the help of the extremal equations
eq.(\ref{eqextre},\ref{eqextre1}), as a function  of $\tan \beta$
and the pseudo-scalar mass $m_{A^0}=\sqrt{m_1^2+m_2^2}$, which
have a more transparent physical meaning. Moreover, we note that
in order to incorporate leading one-loop contributions to the
tree-level potential $V|_{EW}$, eq.(\ref{potEW}), and therefore
trust the results obtained with it up to one-loop level, the
parameters should be evaluated at an appropriate renormalization
scale $Q \sim Q_{SUSY}$, where this SUSY scale is an average of
the typical SUSY masses at the EW vacuum \cite{CCB2,V1Q}. We will
come back to this particular point in sec.4.\\
 Comparison of the depth of the MSSM potential
at the realistic EW vacuum and at the CCB vacuum induces in
addition a new non-trivial relation with the three remaining free
parameters $A_t, m_{\tilde{t}_L}, m_{\tilde{t}_L}$ which enter the
potential $V_3$, eq.(\ref{V3ch}). As a result, a critical bound
$A_{t,3}^c$ above which CCB occurs is identified:
\begin{eqnarray}
\label{CCBcond}
 CCB & \Leftrightarrow& <V_3> \ \ \ \ < \ \ \ \ <V>|_{EW}
\\ & \Leftrightarrow & A_t
> A_{t,3}^c[m_{\tilde{t}_L},
m_{\tilde{t}_R};m_1,m_2,m_3,Y_t,g_1,g_2,g_3]
\end{eqnarray}
We anticipate again on sec.4 and stress that this comparison of
the depth of the tree-level potential at both vacua, and
ultimately the value of the critical bound $A_{t,3}^c$, also
incorporates leading one-loop contributions, provided all
parameters are evaluated at the renormalization scale $Q \sim
Q_{SUSY}$. To investigate this point and determine $A_{t,3}^c$, we
need obviously a precise knowledge on the location and geometry of
the CCB vacua. Sec. 3 is devoted to this particular topic. For a
rapid overview of the situation the interested reader may also
refer to sec.5 where we summarize some important points of this
derivation and detail the practical steps to obtain the critical
bound $A_{t,3}^c$.

\section{The CCB vacuum in the plane $(H_2, \tilde{t}_L, \tilde{t}_R)$}

\subsection{The algorithm to compute the CCB VEVs}

To evaluate the CCB VEVs $<H_2>$ and $<f>$, a numerical step is
now required. A numerical algorithm can be used for instance to
solve simultaneously the two extremal equations
eqs.(\ref{eqH23},\ref{eqf3}). Such a method is however unable to
bring any precise analytical information on the simple geometric
behaviour of the CCB extrema of the potential $V_3$,
eq.(\ref{V3ch}). Alternatively, we propose a procedure which has
the good numerical properties of being fast, secure and easily
implementable on a computer. Moreover, excellent analytical
approximations for the CCB VEVs (at the level of the percent),
and, ultimately, for the optimal conditions on $A_t$ to avoid CCB
can be obtained with it. Finally, it can be easily adapted to the
extended planes $(H_1,H_2, \tilde{t}_L, \tilde{t}_R)$ and
$(H_1,H_2, \tilde{t}_L, \tilde{t}_R, \tilde{\nu}_L)$, first
considered in \cite{CCB2}, and this will enable us to shed new
light on vacuum stability in these directions in a fully
model-independent way \cite{CCB4champs,CCB5champs}.\\
 This alternative procedure to evaluate the CCB VEVs may be
summarized as follows: we first insert an initial value $f^{(0)}$
in the extremal equation associated with $H_2$, eq.(\ref{eqH23}).
This equation is then solved in $H_2$. A solution $H_2^{(0)}$,
which proves to be close to the CCB VEV $< H_2>$, is found. This
solution is then inserted in the extremal equation associated with
$f$, eq.(\ref{eqf3}), which is solved in $f$. We obtain an
improved value $f^{(1)}$, closer to $<f>$ than $f^{(0)}$. The
method is then iterated in a similar way. As a result, we obtain a
set of numerical values $(H_2^{(n)},f^{(n)})_{n \ge 0}$ which
proves to converge fast toward the true CCB extremal set
$(<H_2>,<f>)$. \\
 More precisely, geometrical considerations
confirmed by numerical analysis show that for a given set of free
parameters $A_t,m_{\tilde{t}_L}, m_{\tilde{t}_R}, ...$ the
potential $V_3$, eq.(\ref{V3ch}), may have only two non-trivial
extrema with $<\tilde{t}_{L,R}> \neq 0$: a CCB local minimum and a
CCB saddle-point. Numerical analysis thus splits into two distinct
branches, each one concerning one extremum. For simplicity, in
this article we will not consider the behaviour of the CCB
saddle-point solution, which is useful essentially for
metastability considerations \cite{CCB5,CCB6}. Concerning the
local CCB minimum, the apparent ambiguity in the implementation of
the algorithm on the correct solutions $(H_2^{(n)}, f^{(n)})$ to
choose for each value of  $n \ge 0$ is easily lifted. As will be
shown in sec.3.3, when a CCB minimum develops, the extremal
equation associated with $H_2$, eq.(\ref{eqH23}), has necessarily
three real positive roots in $H_2$. The correct solution
$(H_2^{(n)})_{n \ge 0}$ to follow is always the intermediate one.
There is also no real ambiguity in the choice of $(f^{(n)})_{n \ge
1}$, because the extremal equation associated with $f$,
eq.(\ref{eqf3}), has always only one real positive root in $f$,
which is our candidate. Besides, the solutions of the extremal
equation associated with $H_2$, eq.(\ref{eqH23}) [which gives the
set of values $(H_2^{(n)})_{n \ge 0}$], prove to vary very slowly
as a function of $f$. This feature tends to boost the convergence
of the procedure. Actually, starting with a clever choice for the
input value $f^{(0)}$, the convergence is accelerated so that only
one iteration is needed to fit the exact result with a precision
of $1 \% $ or less, providing ultimately accurate analytical
approximations for the CCB VEVs.\\
 For completeness, let us briefly compare this method to evaluate the VEVs
with the one presented in ref.\cite{CCB2}. Assuming alignment of
the CCB vacuum in the $SU(3)_c$ D-flat direction, as done in
\cite{CCB2}, i.e. $<f>=1$, we obtain easily analytical expressions
for the CCB VEVs depending only on the free parameters
$A_t,m_{\tilde{t}_L}, m_{\tilde{t}_R}, ...$, whereas with the
method presented in \cite{CCB2} a numerical scan is still
required: taking $f^{(0)}=<f>=1$, the VEV $<H_2>$ is simply
obtained with our method by solving analytically the cubic
extremal equation eq.(\ref{eqH23}); the squark fields VEVs
$<\tilde{t}_L>=<\tilde{t}_R>$ are finally obtained by
eq.(\ref{eqUL}). In addition, our iterative algorithm enables us
to take into account, with any desired accuracy, any deviation of
the CCB vacuum from the $SU(3)_c$ D-flat direction. As noted
before, in a model-independent way, such a deviation typically
occurs. This point will be investigated more attentively in the
next section, sec.3.2,  by considering the extremal equation
associated with $f$, eq.(\ref{eqf3}). A subsequent study of the
extremal equation associated with $H_2$, eq.(\ref{eqH23}), will
also provide us with a model-independent optimal bound on $A_t$,
above which a local CCB vacuum begins to develop in the plane
$(H_2,\tilde{t}_L,\tilde{t}_R)$. This point will be addressed in
sec.3.3.

\subsection{The deviation from the $SU(3)_c$ D-flat direction}

We consider now deviations of the CCB vacuum from the $SU(3)_c$
D-flat direction. In ref.\cite{CCB2}, it was argued that, in a
model-independent way, the CCB vacuum is located very close to
this D-flat direction. As noted before, we disagree with this
statement and show in this section that this assumption is
model-dependent. Actually, such a deviation can be quite large, in
particular for large discrepancies between the soft squark masses
$m_{\tilde{t}_L}, m_{\tilde{t}_R}$, and results in a substantial
enhancement of the necessary and sufficient condition to avoid
CCB, $A_t \le A_{t,3}^c$, eq.(\ref{CCBcond}). In fact, the
critical bound $A_{t,3}^c$ can be shown to become more restrictive
and this feature is essential, on a phenomenological ground, when
it comes to relate CCB conditions with the requirement of avoiding
a tachyonic lightest stop. As is well-known, a too large trilinear
soft term $A_t$ can increase this danger, but also any discrepancy
$m_{\tilde{t}_L} \neq m_{\tilde{t}_R}$. The latter effect can be
compensated only by taking into account the deviation of the CCB
vacuum from the $SU(3)_c$ D-flat direction.\\
 The parameter
controlling the deviation of the CCB vacuum from the $SU(3)_c$
D-flat direction is the VEV $<f>$. In the framework of our
algorithm to compute the CCB VEVs, an educated guess for the
initial value $f^{(0)}$ should incorporate information on the
extremal equation associated with $f$, eq.(\ref{eqf3}). Numerical
analysis shows that, to an excellent accuracy, $<f>$ is related to
$<H_2>$ by the relation
\begin{equation}
\label{approxf}
 <f>\sim
\bar{f}(<H_2>)\equiv\sqrt{{\frac{m_{\tilde{t}_L}^2+ Y_t^2
<H_{2}>^2}{m_{\tilde{t}_R}^2+Y_t^2 <H_{2}>^2}}}
\end{equation}
If we neglect gauge contributions, $\bar{f}(H_2)$ is actually the
exact solution to eq.({\ref{eqf3}}). Therefore, this numerical
observation simply reflects the fact that the deviation of the CCB
vacuum from the $SU(3)_c$ D-flat direction is nearly independent
of the D-terms contributions in the potential $V_3$,
eq.(\ref{V3ch}). \\
\begin{figure}[htb]
\vspace*{-0.8cm}
\begin{center}
\mbox{ \psfig{figure=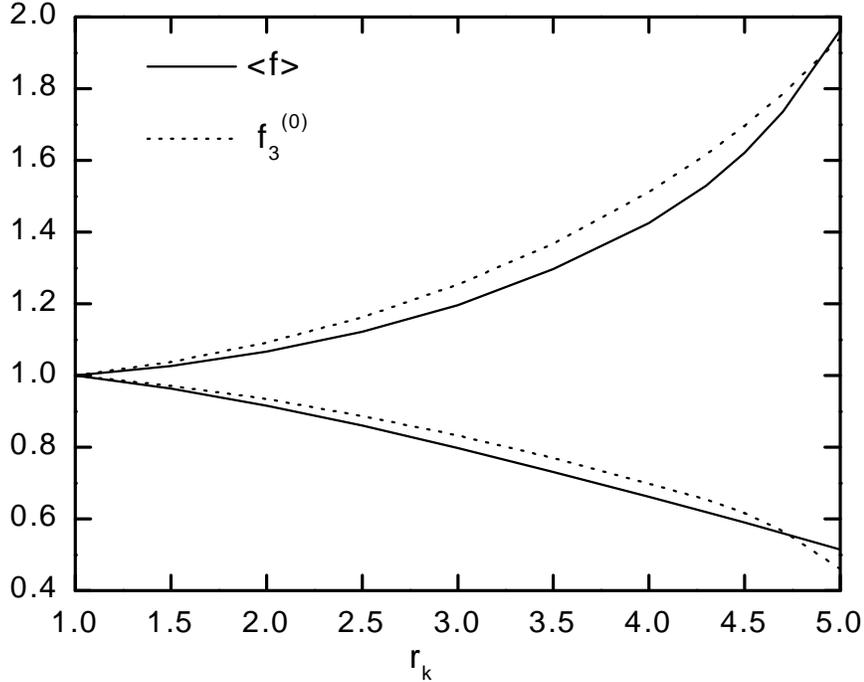,width=13cm}} \vspace*{-1.8cm}
\end{center}
\caption[]{$<f>$, $f^{(0)}_3$ versus
$r_k\equiv(m_{\tilde{t}_L}/{m_{\tilde{t}_R}})^k$, for $k= 1$, $
m_{\tilde{t}_R}=200 \ GeV$  (upper curves) and $k=-1$, $
m_{\tilde{t}_L}=200 \ GeV$ (lower curves). We take $A_t=1400 \
GeV, \ m_2=118 \ GeV, \ Y_t= 1.005, \ g_1=0.356, \ g_2=0.649, \
g_3=1.14$.}
\end{figure}
 To go further, we need to approximate $<H_2>$.
We note first that $\bar{f}(H_2)$ is a slowly varying function of
$H_2$, so that this approximation does not need to be very
accurate. Neglecting in the potential $V_3$, eq.(\ref{V3ch}), the
contributions of the gauge terms and the Higgs mass term $m_2^2$,
and writing the potential as a function of $B_3$, eq.(\ref{B3}),
we find $V_3 \sim-B_3^2/(16 Y_t^2 f^2)$. This simple expression
shows that a rough estimate of the VEV $<H_2>$ is given by the
minimal value taken by $B_3$, eq.(\ref{B3}):
\begin{equation}
 \label{H2ap}
  <H_2> \sim {\frac{A_t}{Y_t}} {\frac{ <f>}{(1+<f>^2)}}
\end{equation}
In fact, the exact VEV $<H_2>$ is numerically typically found to
be lower than this approximate value, but the latter already
contains useful enough information for our purpose. Taking this
value and solving $<f>=\bar{f} (<H_2>)$, we obtain in turn the
excellent approximation to $<f>$
\begin{equation}
\label{f0ap3} <f> \sim f^{(0)}_3 \equiv \sqrt{{\frac{A_t^2+2
m_{\tilde{t}_L}^2-m_{\tilde{t}_R}^2}{A_t^2+2
m_{\tilde{t}_R}^2-m_{\tilde{t}_L}^2}}}
\end{equation}
This approximate value $f^{(0)}_{3}$  is equal to 1 (alignment in
the $SU(3)_c$ D-flat direction) for
$m_{\tilde{t}_L}^2=m_{\tilde{t}_R}^2$. This correctly reproduces
the expected behaviour for $<f>$: when the mass relation
$m_{\tilde{t}_L}^2=m_{\tilde{t}_R}^2$ holds, the potential $V_3$,
eq.(\ref{V3ch}), has an underlying approximate symmetry
$\tilde{t}_L \leftrightarrow \tilde{t}_R$ broken by tiny
$O(g_1^2,g_2^2)$ contributions, so that any non-trivial extremum
must be nearly aligned in the $SU(3)_c$ D-flat direction. In the
large $A_t$ regime, we have also $f^{(0)}_{3} \rightarrow 1$,
reproducing again the expected behaviour for $<f>$. Quite
similarly to the small Yukawa coupling regime, in this limit, the
VEVs of the CCB vacuum become very large. The vacuum then moves
towards the $SU(3)_c \times SU(2)_L \times U(1)_Y$ D-flat
direction, any splitting between the soft squark masses becoming
inessential.\\
 In Figure 1, we illustrate the evolution of the
exact VEV $<f>$ and the approximation $f^{(0)}_3$,
eq.(\ref{f0ap3}), as a function of the ratio of the soft squark
masses $r_k \equiv (m_{\tilde{t}_L}/{m_{\tilde{t}_R}})^k$ for two
cases $k=\pm 1$. The exact VEV $<f>$ has been computed with the
recursive algorithm presented above [see also sec.5 for a
practical summary], taking for initial value $f^{(0)}=f^{(0)}_3$,
eq.(\ref{f0ap3}). In both cases, we have taken $A_t=1400 \ GeV$.
This implies in particular that the expression $f_3^{(0)}$,
eq.(\ref{f0ap3}), is defined up to $r_k\sim 7.14$. However, this
approximation is appropriate only if a CCB vacuum exists. The
sufficient bound given by eq.(\ref{condsuf}) already shows that a
CCB vacuum can develop only for $r_k \le 6$. Once optimized, the
sufficient bound to avoid CCB restricts even more the allowed
range to $r_k \lsim 5.3$ [see e.g. eq.(\ref{csufap}) in sec.3.3].
In Fig.1, the evolution of the VEV is stopped at the boundary
value $r_k \sim 5$, because the CCB vacuum becomes dangerous and
deeper than the EW vacuum only below this value.\\
 The first prominent feature of this
illustration is that $f^{(0)}_3$ is an excellent approximation:
$f^{(0)}_3$ fits $<f>$ with a precision of order $5 \%$, or even
better in the vicinity of $r_k \sim 1$. Moreover, we note that for
a large splitting of the soft squark masses, the deviation of the
CCB global minimum from the $SU(3)_c$ D-flat direction can be
quite large, in particular in the vicinity of the critical bound
$A_{t,3}^c$, i.e. for $r_k \sim 5$. It is smaller for $r_k \sim
1$, where the CCB vacuum is nearly aligned in the $SU(3)_c$ D-flat
direction.\\
 Finally, we may derive refined bounds on the
deviation of the CCB vacuum from the $SU(3)_c$ D-flat direction by
combining the sufficient bound $A_t^{(0)}$, eq.(\ref{condsuf}),
with the accurate approximation $f_3^{(0)}$, eq.(\ref{f0ap3}).
Requiring $A_t \ge A_t^{(0)} \sim m_{\tilde{t}_L}(1+r_1)$, we
obtain:
\begin{equation}
\label{fmin}
 For \ r_1 \equiv {\frac{m_{\tilde{t}_L}}{m_{\tilde{t}_R}}} \ge 1
\ , \ \ \ \ \ \ \ \ \ 1 \le \ <f> \sim f_3^{(0)} \ \ \ \lsim \
\sqrt{{\frac{r_1 (3 r_1+2)}{2 r_1+3}}}
 \end{equation}
For $r_1 \le 1$ these inequalities are reversed.\\
 In an mSUGRA scenario \cite{SOFT,mSUGRA}, we fall typically in the
regime $r_1 \ge 1$, with furthermore $r_1$ perturbatively close to
1. Eq.(\ref{fmin}) then implies $ 1 \le <f> \lsim 1+{\frac{3}{5}}
(r_1-1)$, showing that the CCB vacuum is indeed located in the
vicinity of the $SU(3)_c$ D-flat direction. Such a feature was
built-in through the procedure proposed to evaluate the CCB
conditions in ref.\cite{CCB2}, quite consistently with the mSUGRA
numerical illustration presented in this article. However, it is
important to stress that this assumption is model-dependent, and
may be badly violated in other circumstances near the critical
value $A_{t,3}^c$, in particular in scenarii incorporating
non-universalities of the soft squark masses where the splitting
parameter $r_1$ can be rather large \cite{anom}.

\subsection{The optimal sufficient bound on $A_t$ to avoid CCB}

We consider now more attentively the extremal equation associated
with $H_2$, eq.(\ref{eqH23}). This complementary equation will in
fact enable us to improve the sufficient bound $A_t^{(0)}$,
eq.(\ref{condsuf}), to avoid a dangerous CCB vacuum in the plane
$(H_2, \tilde{t}_L, \tilde{t}_R)$. From a geometrical point of
view, it is reasonable to define the optimal sufficient CCB bound
on $A_t$ to be the largest value below which a local CCB minimum,
not necessarily global, cannot develop. Equivalently, this bound,
denoted $A_t^{suf}$ in the following, is also the critical value
above which a local CCB vacuum begins to develop in the plane
$(H_2, \tilde{t}_L, \tilde{t}_R)$.\\
 The determination of the optimal sufficient bound $A_t^{suf}$
simply requires some additional pieces of information on the
extremal equation associated with $H_2$, eq.(\ref{eqH23}), and on
the geometry of potential $V_3$, eq.(\ref{V3ch}). In order not to
surcharge the text, we will not enter here in the details of this
derivation, but rather refer the reader to the Appendix A. To
summarize, on the technical side, the essential result we obtain
is that if the extremal equation, eq.(\ref{eqH23}), considered as
a cubic polynomial in $H_2$, has only one real root in $H_2$ for
any given value of $f$, then necessarily no local CCB minimum in
the plane $(H_2, \tilde{t}_L, \tilde{t}_R)$ can develop. More
intuitively, this result merely reflects the fact that if a local
CCB vacuum develops with non trivial VEVs $(<H_2>, <f>)$, then on
any path connecting the local extremum at the origin of the fields
to this CCB vacuum, there will be necessarily a saddle on the top
of the barrier separating them, with $\tilde{t}_R, \tilde{t}_L
\neq 0$. For $f=<f>$, such a point will in turn necessarily
correspond to a second real solution in $H_2$ for the extremal
equation, eq.(\ref{eqH23}), in contradiction with the initial
assumption.\\
 Considering the extremal equation
eq.(\ref{eqH23}) as a cubic polynomial in $H_2$, a necessary and
sufficient condition to have only one real root is
\begin{equation}
\label{condsuf1}
 {\cal{C}}_3\equiv [2 \beta_3^3-9
\alpha_3 \beta_3 \gamma_3+ 27 \alpha_3^2 \delta_3]^2+4
[-\beta_3^2+3 \alpha_3 \gamma_3]^3 \ge 0
\end{equation}
The next step to evaluate the optimal sufficient bound $A_t^{suf}$
is to consider this complicated inequality in the direction of a
possible CCB minimum $f= \ <f>$. Taking instead the approximate
value $f \sim f^{(0)}_3$, given by eq.(\ref{f0ap3}), and taking
also values for all the parameters except the trilinear soft term
$A_t$, the equation ${\cal{C}}_3=0$ may be solved numerically as a
function of $A_t$ \footnote{Comparing with a more accurate value
for $<f>$, we found that the maximal discrepancy between the
results obtained is negligible, less than $1 \ GeV$.}. Let us
denote $A_t^{(1)}$ the largest solution of this equation. For $A_t
\le A_t^{(1)}$, we find numerically that we always have
 ${\cal{C}}_3 \ge 0$, showing that there can be no CCB vacuum in
this case. Moreover, numerical investigation also shows that
$A_t^{(1)}$ has typically the desired property of being larger
than $A_t^{(0)}$, eq.(\ref{condsuf}), and, therefore, improves
this bound. There is only one exception to this statement, which
occurs for $m_2^2 \le 0$ and $m_{\tilde{t}_L},m_{\tilde{t}_R} \sim
m_t$. As will be explained in the next section sec.3.4, this
regime actually corresponds to a rather particular situation where
no dangerous CCB vacuum deeper that the EW vacuum may develop,
unless the EW vacuum is unstable.\\
 Taking into account this observation, numerical investigation
finally shows that for $A_t \ge Max [A_t^{(0)}, A_t^{(1)}]$, a
local CCB vacuum begins to develop in the plane $(H_2,\tilde{t}_L,
\tilde{t}_R)$. Hence, this critical value fulfills the properties
required to be identified with the optimal sufficient bound
$A_{t}^{suf}$. In conclusion, we may write without loss of
generality:
\begin{equation}
 \label{newsuf}
A_t \le  A_{t}^{suf} \equiv Max [A_t^{(0)}, A_t^{(1)}]
\Leftrightarrow \ No \ local \ CCB \ vacuum
\end{equation}
where $A_t^{(0)}$ is given by eq.(\ref{condsuf}) and $A_t^{(1)}$
is obtained by solving ${\cal{C}}_3=0$, eq.(\ref{condsuf1}), as
mentioned above. It is important to stress here that this optimal
sufficient bound, obtained with exact analytical expressions,
incorporates all possible deviations of the CCB local vacuum from
the D-flat directions, including the $SU(3)_c$ D-flat one.\\
 As $A_t$ increases above $A_{t}^{suf}$,
the CCB local vacuum soon becomes global and deeper than the EW
vacuum. Obviously, the critical bound $A_{t,3}^c$,
eq.(\ref{CCBcond}), is necessarily larger than $A_t^{suf}$:
\begin{equation}
\label{compsuf}
 A_{t,3}^c \ge A_t^{suf}
\end{equation}
Moreover, we expect that the critical bound $A_{t,3}^c $ should be
perturbatively close to the optimal sufficient bound
$A_{t}^{suf}$, i.e. $A_{t,3}^c \gsim A_{t}^{suf}$, because the EW
potential is not very deep, $<V>|_{EW} \sim -m_Z^4/(g_1^2+g_2^2)$.
Indeed, as will be illustrated in sec.5, the critical bound
$A_{t,3}^c$ is typically located in a range of $5 \%$ or less
above $A_{t}^{suf}$. This interesting feature will considerably
simplify the exact determination of $A_{t,3}^c$, which will be
simply obtained by scanning a small interval in $A_t$ above
$A_{t}^{suf}$. We note finally that in the interesting
phenomenological regime $m_{\tilde{t}_L},m_{\tilde{t}_R} \gsim 300
\ GeV$, a simple empirical approximation of $A_{t}^{suf}$ may be
obtained numerically. We find on one hand $A_{t}^{suf}=A_t^{(1)}$,
with furthermore:
\begin{equation}
\label{csufap} A_t^{suf}=A_t^{(1)} \sim A_t^{ap}\equiv
m_{\tilde{t}_L}+m_{\tilde{t}_R}+|m_2|
\end{equation}
This approximation exhibits in which amount the sufficient bound
$A_t^{(0)}$ is improved in this regime. The difference is of order
$|m_2|$: $A_t^{(1)}-A_t^{(0)} \sim |m_2|$.\\

Let us come back  briefly now to the implementation of the
procedure to compute the CCB VEVs. We have shown that for $A_t \le
A_t^{suf}$, eq.(\ref{newsuf}), no local CCB vacuum may develop in
the plane $(H_2,\tilde{t}_L, \tilde{t}_R)$. As noted before, in
the dangerous complementary regime, $A_t \ge A_t^{suf}$, we need
to evaluate the VEVs of the CCB local vacuum in order to compare
the depth of the CCB potential and the EW potential and find the
necessary and sufficient bound $A_{t,3}^c$, eq.(\ref{CCBcond}), to
avoid CCB. For $f=f^{(0)}_3$, eq.(\ref{f0ap3}), the extremal
equation associated with $H_2$, eq.(\ref{eqH23}), has three real
roots, which also prove to be positive [see Appendix A]. The
intermediate root, denoted $H_2^{(0)}$ in sec.3.1, proves to be an
excellent approximation of the VEV $<H_2>$ [at a level of \lsim 1
\%] \footnote{For completeness, we note that typically the lowest
solution will correspond to a directional CCB saddle-point,
whereas the largest is spurious, giving $<\tilde{t}_L>^2 \le 0$}.
The analytic expression of $H_2^{(0)}$ is complicated and not
particularly telling, therefore we refrain from giving it here.
This shows that, to an excellent approximation, we can obtain
explicit analytic expressions for all the CCB VEVs: $(<H_2>, <f>)$
are approximated by $(H_2^{(0)},f^{(0)}_3)$, where $f^{(0)}_3$
enables us to take into account the deviation of the CCB vacuum
from the $SU(3)_c$ D-flat direction, and the squark VEVs
$<\tilde{t}_{L/R}>$ are subsequently obtained by
eqs.(\ref{ft},\ref{eqUL}). This way, we can obtain in turn an
accurate analytical expression for the CCB potential $V_3$,
eq.(\ref{V3ch}), at the CCB vacuum. Comparison with the EW
potential $<V>|_{EW}$, eq.(\ref{VEW}), ultimately provides an
excellent approximation of the critical CCB bound $A_{t,3}^c$,
eq.(\ref{CCBcond}). The accuracy of this approximation can be
improved at will  by iterating the procedure to compute the CCB
VEVs, as depicted in sec.3.1. We note however that the impact on
$A_{t,3}^c$ is negligible, $ \sim O(1 \ GeV)$.

\subsection{The instability condition of the EW potential}

Besides the contribution of the trilinear soft term $A_t$, another
negative contribution in the potential $V_3$, eq.(\ref{V3ch}),
appears at the EW scale when the Higgs parameter $m_2^2$ becomes
negative. At the tree-level, the sign of $m_2^2$ is related in a
simple way to $\tan \beta$ by the extremal equation
eq.(\ref{eqextre}) in the EW direction: for $\tan \beta \ge
\sqrt{1+2 m_{A^0}^2/m_{Z^0}^2}$, $m_2^2$ is negative, whereas it
is positive in the complementary low $\tan \beta$ regime.
Numerical investigation shows that we have only two distinct
patterns for the number of local extrema of $V_3$,
eq.(\ref{V3ch}). They are distinguished by the sign of $m_2^2$
and, consequently, the magnitude of $\tan \beta$.\\

{\bf $\bullet$ $m_2^2 \ge 0:$} \\

The potential $V_3$, eq.(\ref{V3ch}), has one trivial local
minimum, namely the origin of the fields, and possibly a pair of
non-trivial local extrema with $<\tilde{t}_{L,R}> \neq 0$: the
would-be global CCB vacuum and a CCB saddle-point sitting on top
of the barrier separating it from the origin of the fields. In
this case, the optimal sufficient bound $A_{t}^{suf}$,
eq.(\ref{newsuf}), always proves to be equal to $A_t^{(1)}$.
Besides, for soft squark masses large enough, i.e.
$m_{\tilde{t}_L}, m_{\tilde{t}_R} \gsim |m_2|$, the traditional
CCB bound in the D-flat direction $A_{t,3}^D$,
eq.(\ref{condfrere}), typically provides an upper bound for the
critical bound $A_{t,3}^c$, eq.(\ref{CCBcond}), so that we may
write:
\begin{equation}
\label{relacm2p}
 A_{t}^{suf}=A_t^{(1)} \le A_{t,3}^c \le A_{t,3}^D
\end{equation}
We note however that this upper bound is not very indicative of
the critical value $A_{t,3}^c$ for large values of the soft
masses, $m_{\tilde{t}_L}, m_{\tilde{t}_R} \gg |m_2|$, as will be
illustrated in sec.5. Actually, in this regime, the relation
eq.(\ref{massrel}) which is the signature of an alignment in the
D-flat direction is badly violated, implying a large deviation of
the CCB vacuum from the D-flat direction.\\

{\bf $\bullet$ $m_2^2 \le 0:$} \\

Besides the origin of the fields and possibly a pair of CCB
extrema (a local minimum and a saddle-point), the potential $V_3$,
eq.(\ref{V3ch}), has another non-trivial extremum with VEVs
$<H_2>_{\overline{EW}}^2 = -4 m_2^2/(g_1^2+g_2^2)$, $
<\tilde{t}_{R,L}>_{\overline{EW}}=0$, giving
$<V_3>_{\overline{EW}}=-2 m_2^4/(g_1^2+g_2^2)$. The origin of the
fields is now unstable and the potential automatically bends down
in the direction of this non-CCB extremum. We note also that, for
large $\tan \beta$, the EW vacuum tends towards it as the inverse
power of $\tan \beta$: $(v_1=(174 GeV)/\sqrt{1+\tan^2 \beta},v_2)
\rightarrow (0,$ $ <H_2>_{\overline{EW}})$. [Accordingly, the
negativity of $m_2^2$ appears as a mark in the plane
$(H_2,\tilde{t}_L,\tilde{t}_R)$ of the well-known instability
condition at the origin of the fields $m_1^2 m_2^2 -m_3^4 \le 0$,
which is the signal of an EW symmetry breaking \cite{MSSM}].\\
 Obviously, if this additional extremum is a saddle-point of
the potential $V_3$, eq.(\ref{V3ch}), then a deeper CCB minimum is
necessarily present in the plane $(H_2,\tilde{t}_L,\tilde{t}_R)$.
The squared mass matrix evaluated at the non-CCB extremum reads
 \begin{eqnarray}
 \label{matst1}
 {\cal M}^2|_{\overline{EW}}=\left(\begin{array}{ccc}
{\frac{g_1^2+g_2^2}{2}} H_2^2& 0 & 0
\\ [0.4cm] 0  &  m_{\tilde{t}_L}^2 +Y_t^2
 H_2^2(1-{\frac{(3 g_2^2-g_1^2)}{12 Y_t^2}})
  & -A_t Y_t H_2
\\ [0.4cm] 0 & -A_t Y_t H_2 &
 m_{\tilde{t}_R}^2+ Y_t^2
 H_2^2 (1-{\frac{g_1^2}{3 Y_t^2}})
\end{array} \right)\\ [0.4cm]
\nonumber
\end{eqnarray}
with $H_2=<H_2>_{\overline{EW}}$.\\
 Stability of the non-CCB
vacuum is equivalent to the positivity of all the squared mass
eigenvalues of ${\cal M}^2|_{\overline{EW}}$, eq.(\ref{matst1}).
It is not automatic and needs
\begin{equation}
\label{Ameta1}
 A_t \le A_t^{inst}
\end{equation}
where
\begin{eqnarray}
\label{Ameta2}
 (A_t^{inst})^2 &\equiv& m_{\tilde{t}_L}^2
(1-{\frac{g_1^2}{3 Y_t^2}})+m_{\tilde{t}_R}^2 (1-{\frac{(3
g_2^2-g_1^2)}{12 Y_t^2}})-{\frac{g_1^2+g_2^2}{4 m_2^2}}
{\frac{m_{\tilde{t}_L}^2 m_{\tilde{t}_R}^2}{Y_t^2}}
\\ \nonumber
 &&-{\frac{4 m_2^2}{g_1^2+g_2^2}} Y_t^2(1-{\frac{(3
g_2^2-g_1^2)}{12 Y_t^2}}) (1-{\frac{g_1^2}{3 Y_t^2}})
\end{eqnarray}
Let us remark that, for $\tan \beta \rightarrow + \infty$, the
lower $2 \times 2$ matrix of ${\cal M}^2|_{\overline{EW}}$,
eq.(\ref{matst1}), is simply equal to the tree-level physical
squared stop mass matrix \cite{MSSM}, so that the instability
condition, eq.(\ref{Ameta1}), is a mere rephrasing of the physical
requirement of avoiding a tachyonic lightest stop, expressed as a
function of $A_t$. This statement is also valid to a good accuracy
when the stop mixing parameter $\tilde{A}_t=A_t+ \mu/ \tan \beta$
is well approximated by the trilinear soft term $A_t$, i.e. for
$|\mu| \ll |A_t| \tan \beta$.\\
 To simplify the discussion, in the following we will essentially
identify this non-CCB extremum with the EW vacuum, implying in
particular that the potential at both vacua are equal, i.e.
$<V>|_{EW}$ $ \sim <V_3>_{\overline{EW}}$. This assumption,
accurate for $\tan \beta$ large enough, enables us to write the
following relation on the CCB bounds
\begin{equation}
\label{relac}
 A_t^{suf} \le A_{t,3}^c \le A_t^{inst}
\end{equation}
The first relation was actually obtained in the last section, see
eq.(\ref{compsuf}), whereas the second means that if the non-CCB
extremum is unstable, then a dangerous CCB vacuum, deeper than the
EW vacuum\footnote{The relation $A_{t,3}^c \le A_t^{inst}$ is
still accurate if we relax our simplifying assumption $<V>|_{EW}
\sim$ $<V_3>_{\overline{EW}}$, because the potential $V_3$ deepens
rapidly with increasing $A_t$.}, has developed in the plane
$(H_2,\tilde{t}_L,\tilde{t}_R)$.\\
 For $m_2^2 \le 0$, the relation (\ref{relac}) provides
the most general upper and lower bounds on the critical CCB bound
$A_{t,3}^c$, eq.(\ref{CCBcond}). We note however that in the
interesting regime of large squark soft masses, i.e.
$m_{\tilde{t}_L}, m_{\tilde{t}_R} \ge m_t$, the upper bound given
by $A_t^{inst}$ is typically largely improved by the traditional
bound in the D-flat direction $A_{t,3}^D$, eq.(\ref{condfrere}).
However, quite similarly to the case $m_2^2 \ge 0$, the latter
bound $A_{t,3}^D$ is itself typically very large compared to the
critical CCB bound $A_{t,3}^c$, due to a large deviation of the
CCB vacuum from the D-flat directions. A better indication of the
critical CCB bound $A_{t,3}^c$, eq.(\ref{CCBcond}), is always
given by the optimal sufficient bound $A_t^{suf}$,
eq.(\ref{newsuf}).\\

Finally, let us consider more attentively the behaviour of the
potential in the limit $A_t \rightarrow A_t^{inst}$. This will
enlighten the importance of taking into account any deviation of
the CCB vacuum from the $SU(3)_c$ D-flat direction in order to
obtain a consistent critical CCB bound $A_{t,3}^c$,
eq.(\ref{CCBcond}), which encompasses the possibility of avoiding
a tachyonic stop mass. Two interesting different modes with
particular geometrical features of the potential can be
considered:\\
 {\bf{i)}} The CCB vacuum is located away from the
non-CCB extremum. This possibility in fact corresponds either to
the case $m_{\tilde{t}_L} m_{\tilde{t}_R} \ll m_t^2$ or
$m_{\tilde{t}_L} m_{\tilde{t}_R} \gg m_t^2$, where $m_t$ is the
top quark mass. In the first case, the CCB vacuum proves to be
closer to the origin of the fields than the non-CCB extremum,
whereas in the latter this hierarchy is reversed. In both cases,
the optimal sufficient bound $A_t^{suf}$, eq.(\ref{newsuf}), is
always given by $A_t^{(1)}$. In the limit $A_t \rightarrow
A_t^{inst}$, the CCB saddle-point located on top of the barrier
separating the CCB vacuum and the non-CCB extremum tends towards
the non-CCB extremum and the barrier separating both vacua
eventually disappears.\\
 {\bf{ii)}} The CCB vacuum interferes with
the non-CCB vacuum. For $A_t \rightarrow A_t^{inst}$, this mode
corresponds to a degenerate situation where the CCB local vacuum
and the CCB saddle-point overlap and tend towards the non-CCB
vacuum. This possibility appears clearly by comparing the
instability bound $A_t^{inst}$, eq.(\ref{Ameta1}), with the
sufficient bound $A_t^{(0)}$, eq.(\ref{condsuf}). We have
\begin{equation}
\label{consis}
 (A_t^{inst})^2-(A_t^{(0)})^2=[m_{\tilde{t}_L} m_{\tilde{t}_R}-
 (1-{\frac{(3 g_2^2-g_1^2)}{12 Y_t^2}}) (1-{\frac{g_1^2}{3 Y_t^2}}) Y_t^2
H_2^2]^2 {\frac{1}{Y_t^2 H_2^2}} \ge 0
\end{equation}
with $H_2=<H_2>_{\overline{EW}}$. Combining the last equation with
eq.(\ref{relac}), we obtain:
\begin{equation}
\label{satur} m_{\tilde{t}_L} m_{\tilde{t}_R}=[1-{\frac{(3
g_2^2-g_1^2)}{12 Y_t^2}}] [1-{\frac{g_1^2}{3 Y_t^2}}] \ m_t^2
\Leftrightarrow A_{t,3}^c=A_t^{inst}=A_t^{suf}[=A_t^{(0)}]
\end{equation}
where the EW and the non-CCB vacua have been identified to write
$m_t= Y_t <H_2>_{\overline{EW}}$. The equalities on the right hand
side of the equivalence eq.(\ref{satur}) signal that for this
particular values of the soft squark masses, we are at the center
of a critical regime where the CCB vacuum interferes with EW
vacuum. This critical regime actually extends to a small range in
$m_{\tilde{t}_L}, m_{\tilde{t}_R}$ around this center, and is more
generally characterized by the relation $A_t^{inst}=A_{t,3}^c$,
meaning that no dangerous CCB vacuum, deeper than the EW vacuum,
may develop unless the EW vacuum is unstable. In this region,
there is also typically no room for a CCB vacuum to develop, not
even a local one. This occurs already, e.g., at the center of the
critical regime, where we have $A_t^{inst}=A_t^{suf}=[A_t^{(0)}]$.
As will be illustrated in sec.5 [see Fig.4], this critical regime
includes a small domain around this center where the typical
hierarchy $A_t^{(1)} \ge A_t^{(0)}$ is slightly violated, giving
$A_t^{suf}=A_t^{(0)}$, and which is itself  bordered by a domain
where this hierarchy is respected, giving $A_t^{suf}=A_t^{(1)}$.\\
 We come now more precisely to the relation between the critical
CCB bound $A_{t,3}^c$, eq.(\ref{CCBcond}), and the requirement of
avoiding a tachyonic lightest stop. Obviously, this relation is
crucial in the interference regime $m_{\tilde{t}_L}
m_{\tilde{t}_R} \sim m_t^2$, corresponding to the case ii), where
we have $A_t^{inst}=A_{t,3}^c$. With the help of the extremal
equations, it is a straightforward exercise to show that for $A_t
\rightarrow A_t^{inst}[=A_{t,3}^c]$, the VEVs of the CCB extremum
verify $(<H_2>, <\tilde{t}_{L,R}>) \rightarrow
(<H_2>_{\overline{EW}},0)$, with furthermore:
\begin{equation}
\label{fEW}
 <f> \ \rightarrow \sqrt{{\frac{12(m_{\tilde{t}_L}^2+
Y_t^2 <H_{2}>_{\overline{EW}}^2) +(g_1^2-3 g_2^2)
<H_{2}>_{\overline{EW}}^2 }{12(m_{\tilde{t}_R}^2+Y_t^2
<H_{2}>_{\overline{EW}}^2)-4 g_1^2 <H_{2}>_{\overline{EW}}^2}}}
\end{equation}
This particular direction is, in fact, connected to the direction
of the lightest stop eigenstate. Let us denote $(\overline{t}_1,
\overline{t}_2)$ the stop-like eigenstates of the $2 \times 2$
lower matrix in ${\cal M}^2|_{\overline{EW}}$, eq.(\ref{matst1}),
and $\overline{\theta}$ the mixing angle of the rotation matrix
$\cal{R}$ relating these eigenstates to the VEVs $(<\tilde{t}_L>$,
$<\tilde{t}_R>)$:
\begin{eqnarray}
\label{rot1} \left(\begin{array}{c} \overline{t}_1 \\ [0.4cm]
\overline{t}_2
\end{array} \right) ={\cal{R}} \left(\begin{array}{c}
<\tilde{t}_L> \\ [0.4cm] <\tilde{t}_R>
\end{array} \right) \ \ \ with \ \ \
 {\cal{R}} \equiv \left(\begin{array}{cc} cos\overline{\theta}
& sin\overline{\theta}
\\ [0.4cm] -sin\overline{\theta} &
cos\overline{\theta}
\end{array} \right)\\ [0.4cm]
\nonumber
\end{eqnarray}
As noted before, if we assume that $\tan \beta$ is large enough
and $|\mu| \ll |A_t| \tan \beta$, we may safely identify this
matrix with the physical squared stop matrix, and
$(\overline{t}_1, \overline{t}_2,\overline{\theta})$ with the stop
eigenstates and mixing angle $(\tilde{t}_1,
\tilde{t}_2,\tilde{\theta})$ \cite{MSSM}.\\
  By definition, for $A_t
\rightarrow A_t^{inst}$, the matrix ${\cal M}^2|_{\overline{EW}}$,
eq.(\ref{matst1}), has one zero eigenvalue and the wall separating
the CCB extremum and the non-CCB extremum lowers and eventually
disappears in the direction of the corresponding eigenstate
$\overline{t}_1$. In the basis $(\tilde{t}_L, \tilde{t}_R)$, the
components of this eigenstate read
$\overline{t}_1=(\overline{t}_1^L= cos \overline{\theta},
\overline{t}_1^R= sin \overline{\theta})$ and prove to verify
\begin{equation}
\tan \overline{\theta} \equiv
{\frac{\overline{t}_1^R}{\overline{t}_1^L}}=<f>
\end{equation}
where $<f>$ is given by the limiting value in eq.(\ref{fEW}). This
shows on one hand that, in this critical regime, the stop mixing
angle $\overline{\theta}$ is related in a simple way to the
deviation of the CCB vacuum from the $SU(3)_c$ D-flat direction
and, on the other, that taking into account such a deviation of
the CCB vacuum to evaluate the critical CCB bound $A_{t,3}^c$,
eq.(\ref{CCBcond}), is crucial to avoid a tachyonic lightest stop.

\section{Radiative corrections}

In this section, we discuss the renormalization scale at which the
tree-level necessary and sufficient condition to avoid CCB, $A_t
\le A_{t,3}^c$, eq.(\ref{CCBcond}), should be evaluated in order
to incorporate leading one-loop corrections. As is well-known, on
a general ground, the complete, all order effective potential
$V(\phi)$ is a renormalization group invariant. However, this
property is not shared by the tree-level approximation $V^{(0)}$
which typically depends strongly on the renormalization scale $Q$
at which it is computed \cite{CCB2,V1Q}. A kind of renormalization
group-improved version of the tree-level potential which would
incorporate a resummation of all leading logarithmic contributions
would certainly be more reliable. However, one faces here the
tricky problem of dealing with many mass scales \footnote{Some
attempts have be made in this direction, see \cite{RGI}}. A better
approximation to $V(\phi)$, more stable with respect to the scale
$Q$, is in fact given by the one-level effective potential
($\overline{MS}$ scheme) \cite{CCB2,V1Q}
 \begin{equation}
 V^{(1)}(\phi)=V^{(0)}(\phi)+ \sum_i{{\frac{(-1)^{2s_i}(2s_i+1)}{64 \pi^2}}
 M_i^4(\phi)[Log {\frac{M_i^2(\phi)}{Q^2}}-{\frac{3}{2}}]}
 \end{equation}
where $M_i^2(\phi)$ denotes the tree-level squared mass of the
eigenstate labeled i, of spin $s_i$, in the scalar field direction
$\phi$. The scale $Q$ enters explicitly in the one-loop
correction, but also implicitly in the running of the mass and
coupling parameters.\\
 Obviously, in the field direction $(H_2,\tilde{t}_L,\tilde{t}_R)$
studied in this paper, such a one-loop correction will introduce
very complicated field contributions which will modify the simple
tree-level geometrical picture presented here. However, we may
still have "locally" a good indication of the impact of these
radiative corrections with the help of our tree-level
investigation. As is also well-known, around some scale $Q_0$
which depends on the field direction considered, the predictions
obtained with the tree-level potential $V^{(0)}$ and the one-loop
level potential $V^{(1)}$ approximately coincide \cite{CCB2,V1Q}.
This numerical observation was in fact intensively used, in
particular in the context of CCB studies
\cite{CCB2,CCB3,CCB4,CCB5,CCB6}, precisely in order to use the
relative simplicity of the tree-level potential. This
field-dependent scale $Q_0$ is typically of the order of the most
significant mass present in the field region investigated. This
roughly means that we reduce the multi-scale problem to a
one-scale one, the "most significant mass" meaning a kind of
average of the field-dependent masses which provide the leading
one-loop contributions in the direction of interest
\cite{CCB2,V1Q}.\\
 At the EW vacuum, it has been shown that the
appropriate renormalization scale $Q_{SUSY}$ where the one-loop
corrections to the tree-level potential $V|_{EW}$,
eq.(\ref{potEW}), can be safely neglected is an average of the
typical SUSY masses \cite{CCB2,V1Q}. For instance, for large
$M_{SUSY} \sim m_{\tilde{t}_L} \sim m_{\tilde{t}_R} \gg m_t$, the
tree-level potential receives important radiative corrections
coming from loops of top and stop fields. In this case, $Q_{SUSY}$
is expected to be an average of the top and stop masses, giving
$Q_{SUSY} \sim M_{SUSY}$, whereas for low $M_{SUSY} \lsim m_t$,
this scale is somewhat underestimated and should be raised to a
more typical SUSY mass \cite{CCB2,V1Q}. In this light, we see that
we may trust the results obtained with the tree-level potential
$V|_{EW}$, eq.(\ref{potEW}), in particular the EW VEVs $(v_1,v_2)$
given by eqs.(\ref{eqextre}, \ref{eqextre1}) and the depth of the
EW potential $<V>|_{EW}$, eq.(\ref{VEW}), provided all parameters
entering this potential are evaluated at the appropriate scale $Q
\sim Q_{SUSY}$.\\
 What is now the appropriate scale $Q_{CCB}$ where the results obtained
with the tree-level potential $V_3$, eq.(\ref{V3ch}), incorporate
leading one-loop corrections? At the CCB vacuum, such corrections
are expected to be induced by loops involving masses in the scalar
field direction $(H_2, \tilde{t}_L,\tilde{t}_R)$, in particular
for $m_{\tilde{t}_L} \sim m_{\tilde{t}_R} \gg m_t$ for which these
contributions are enhanced. Accordingly, we estimate $Q_{CCB}$ to
be an average of these masses, more precisely $Q_{CCB} \sim
\sqrt{\frac{<Tr{\cal{M}}^2>|_{CCB}}{3}}$, where:
\begin{eqnarray}
 <Tr{\cal{M}}^2>|_{CCB}&=&{\frac{1}{2}} <{\frac{\partial^2 V_3}
 {\partial H_2^2}}+{\frac{\partial^2 V_3}{\partial \tilde{t}_L^2}}+
 {\frac{\partial^2 V_3}{\partial
 \tilde{t}_R^2}}>|_{CCB}\\
&=& m_{\tilde{t}_L}^2+m_{\tilde{t}_R}^2+Y_t^2 [2
H_2^2+{\tilde{t}_L}^2 (f^2+1)]+A_t Y_t f
{\frac{{\tilde{t}_L}^2}{H_2}} \nonumber \\ &&+ {\frac{g_1^2}{36}}
[9 H_2^2+(44 f^2-1) \tilde{t}_L^2]+{\frac{g_2^2}{4}} (H_2^2+3
{\tilde{t}_L}^2)+{\frac{2 g_3^2}{3}} {\tilde{t}_L}^2 (1+f^2)
\end{eqnarray}
All fields should be evaluated at the CCB vacuum. To derive the
last expression, we have used the extremal equation $\partial
V_3/\partial H_2=0$ to replace the Higgs mass parameter $m_2$. The
VEV $<\tilde{t}_L>$ may also be replaced with the help of the
extremal equation eq.(\ref{eqUL}-\ref{B3}), giving a complicated
expression for $Q_{CCB}$ which depends only on the soft terms
$A_t, m_{\tilde{t}_L}, m_{\tilde{t}_R}$, the gauge and Yukawa
couplings and the CCB VEVs $<H_2>, <f>$. For simplicity, let us
take $M_{SUSY} =m_{\tilde{t}_L}=m_{\tilde{t}_R}$, which gives
$<f>= 1$. Taking furthermore $Y_t,g_3 \sim 1$, neglecting other
gauge couplings and (over-)estimating the VEV $<H_2> \sim A_t/2$
 [see eq.(\ref{H2ap})], we find
\begin{equation}
Q_{CCB}^2\sim {\frac{11 A_t^2-20 M_{SUSY}^2}{18}}
\end{equation}
This scale is meaningful only when a CCB vacuum develops, that is
for $A_t \ge A_t^{suf}$ [see eq.(\ref{newsuf})]. For illustration,
we estimate roughly this lower bound with $A_t^{(0)}$,
eq.(\ref{condsuf}). Taking $A_t \sim 2 \ M_{SUSY}$, we obtain
$Q_{CCB} \sim 1.33 \ M_{SUSY}$. Let us stress here that a refined
evaluation of $Q_{CCB}$, with realistic values for the gauge
couplings, the CCB VEV $<H_2>$ and the optimal sufficient bound
$A_t^{suf}$, would give in fact a value for $Q_{CCB}$ closer to
$M_{SUSY}$. This simple illustration however already provides a
clear indication that $Q_{CCB}$ is typically of order
$M_{SUSY}$.\\
 For $M_{SUSY} \gg m_t$ and $A_t \gsim
A_t^{suf}$, we conclude therefore that we have $Q_{CCB} \sim
Q_{SUSY}$. Obviously, a similar conclusion is expected in the
complementary regime $M_{SUSY} \lsim m_t$: in this case, the CCB
and the EW vacua prove to be close, implying a mass spectrum of
the same order at each vacuum. We note also that this estimation
of $Q_{CCB}$ is in full agreement with the one obtained in
ref.\cite{CCB2} in the extended plane
$(H_1,H_2,\tilde{t}_L,\tilde{t}_R)$. In this article, the scale
$Q_{CCB}$ was estimated to be $\sim Max[Q_{SUSY},g_3 A_t/4 Y_t,
A_t/4]$, which reduces for $Y_t,g_3 \sim 1$ and $A_t \gsim
A_t^{suf} \gsim 2 M_{SUSY}$ to $Q_{CCB} \sim Q_{SUSY}$.\\
 Two important conclusions can be deducted from this result. On one hand,
we see that the optimal sufficient bound $A_t^{suf}$,
eq.(\ref{newsuf}), should be evaluated at $Q_{CCB} \sim Q_{SUSY}$,
in order to minimize the one-loop radiative corrections to the
tree-level potential $V_3$, eq.(\ref{V3ch}). More importantly, we
see that, at this common scale $Q_{CCB} \sim Q_{SUSY}$, it is also
meaningful to compare the tree-level depth of the potential at the
EW vacuum, i.e. $<V>|_{EW}$, eq.(\ref{VEW}), and at the CCB
vacuum, in order to determine the necessary and sufficient
condition $A_t \le A_{t,3}^c$, eq.(\ref{CCBcond}), to avoid CCB in
the plane $(H_2,\tilde{t}_L,\tilde{t}_R)$. This point is a mere
consequence of the fact that the potential at a realistic EW
vacuum is not very deep, as already noted in sec.3.3 [see
eq.(\ref{compsuf})], therefore giving $A_{t,3}^c \gsim A_t^{suf}$.
To summarize, we expect our tree-level refined CCB bounds to be
robust under inclusion of leading one-loop corrections to the
potential, provided they are evaluated at $Q \sim Q_{SUSY}$.
Accordingly, stability of the EW vacuum in the plane
$(H_2,\tilde{t}_L,\tilde{t}_R)$ should be tested in
model-dependent scenarii \cite{SOFT,anom,CCB2} at this scale.

\section{Practical guide to evaluate the CCB conditions}

Let us now collect and summarize the main results we have found.
As mentioned in sec.2.3, the evaluation of the critical bound
$A_{t,3}^c$, eq.(\ref{CCBcond}), above which there is CCB in the
plane $(H_2, \tilde{t}_L, \tilde{t}_R)$ requires the precise
determination of the CCB VEVs and comparison of the potential
$V_3$, eq.(\ref{V3ch}), at the CCB vacuum with the value of the
potential at the EW vacuum $<V>|_{EW}$, eq.(\ref{VEW}). This
comparison is meaningful and incorporates leading one-loop
corrections, provided all parameters are evaluated at the
appropriate renormalization scale $Q \sim Q_{SUSY}$, where
$Q_{SUSY}$ is an average of the typical SUSY masses at a realistic
EW vacuum. This assumption will be implicitly made in the
following. Accordingly, the main practical steps to evaluate
$A_{t,3}^c$ are:\\

{\bf $\bullet$ Evaluation of the depth of the EW potential:} take
a realistic set of values for $g_1,g_2,g_3$ consistent with
experimental data; choose in addition values for $\tan \beta$ and
the pseudo-scalar mass $m_{A^0}^2=m_1^2+m_2^2$. The top mass
$m_t=Y_t v \sin \beta$, with $v=174 \ GeV$, determines the value
of the top Yukawa coupling $Y_t$. Finally, the extremal equations
in the EW direction eqs.(\ref{eqextre},\ref{eqextre1}) determine
the Higgs mass parameters $m_1,m_2,m_3$ and the depth of the
potential at the EW vacuum $<V>|_{EW}$, eq.(\ref{VEW}).\\

{\bf $\bullet$ Evaluation of the CCB optimal sufficient bound:}
choose a set of values for the soft mass parameters
$m_{\tilde{t}_L},m_{\tilde{t}_R}$ and evaluate the optimal
sufficient bound $A_t^{suf}= Max [A_t^{(0)}, A_t^{(1)}]$,
eq.(\ref{newsuf}). This requires the comparison of the quantities
$A_t^{(0)}$ given by eq.(\ref{condsuf}), and $A_t^{(1)}$ given by
the largest solution in $A_t$ of the equation ${\cal{C}}_3=0$,
eq.(\ref{condsuf1}). To evaluate $A_t^{(1)}$, the parameter $f$ of
the departure of the CCB vacuum from the $SU(3)_C$ D-flat
direction should be taken at the excellent approximated value
$f^{(0)}_3$, eq.(\ref{f0ap3}) [see sec.3.2]. The value obtained
$A_t^{suf}$ is the optimal sufficient bound to avoid CCB. This
means that for $A_t=A_t^{suf}$ a CCB local vacuum, not necessarily
global, begins to develop in the plane $(H_2, \tilde{t}_L,
\tilde{t}_R)$, and soon becomes global as $A_t$ increases. This
bound therefore considerably simplifies the determination of the
necessary and sufficient bound $A_{t,3}^c$ to avoid CCB in this
plane. \\
 For $1 \le \tan \beta \le
\sqrt{1+2 m_{A^0}^2/m_{Z^0}^2}$, or equivalently $m_2^2 \ge 0$, we
always have $A_t^{suf}=A_t^{(1)}$ [see sec.3.4]. In the
complementary regime, we have $m_2^2 \le 0$, and an additional
non-CCB vacuum develops in the plane $(H_2, \tilde{t}_L,
\tilde{t}_R)$. For large enough $\tan \beta$, it essentially
coincides with the EW vacuum which is located in the vicinity of
this plane [see sec.3.4]. As a result, a new computable
instability bound $A_t^{inst}$, eqs.(\ref{Ameta1},\ref{Ameta2}),
appears. This bound merely reflects the physical requirement of
avoiding a non-tachyonic lightest stop, for large $\tan \beta$.
Besides, an inversion of the typical hierarchy between the
sufficient bounds $A_t^{(0)} \le A_t^{(1)}$ may occur, implying
$A_t^{suf}=A_t^{(0)}$. This inversion however takes place only in
the critical region $m_{\tilde{t}_L} m_{\tilde{t}_R} \sim m_t$
where the CCB vacuum interferes with the non-CCB vacuum
aforementioned. In this interference regime, the instability bound
$A_t^{inst}$ is quite restrictive and the relation $A_t^{suf} \le
A_{t,3}^c \le A_t^{inst}$, eq.(\ref{relac}), is saturated on both
sides, meaning that no dangerous CCB vacuum may develop unless the
EW vacuum is unstable.\\
 Whatever the value of $\tan \beta$ is, for large enough soft masses
$m_{\tilde{t}_L}, m_{\tilde{t}_R}\gsim 300 \ GeV$, a good
approximation to the bound $A_t^{suf}$ is given by $A_t^{ap}$,
eq.(\ref{csufap}) [see sec.3.3].\\

Let us remark that the parameters involved in these first two
steps, basically $(m_{A^0},\tan \beta,$ $m_{\tilde{t}_L},
m_{\tilde{t}_R})$, are typical of phenomenological
model-independent Higgs studies, the benchmark scenario
$M_{SUSY}=m_{\tilde{t}_L}=m_{\tilde{t}_R}$ being often considered
\cite{higgs,higgsrev}. Once such a set of values is chosen, CCB
considerations induce an additional constraint on the allowed
values for the trilinear soft term $A_t$.\\

{\bf $\bullet$ Evaluation of the CCB critical bound $A_{t,3}^c$:}
the determination of the CCB VEVs and comparison of the depth of
the CCB potential $V_3$, eq.(\ref{V3ch}), and $<V>|_{EW}$,
eq.(\ref{VEW}), is needed [see sec.2.3]. This step requires a
numerical scan of the region $A_t \ge A_t^{suf}$, which is not
time consuming, because typically the critical bound $A_{t,3}^c$
proves to be just slightly above the optimal sufficient bound
$A_t^{suf}$ previously determined. The computation of the CCB VEVs
may be achieved with the help of the algorithm presented in
sec.3.1. The main steps are the following:\\
  - Solve the extremal equation eq.(\ref{eqH23}) in $H_2$
with the initial input $f=f^{(0)}_3$ given by eq.(\ref{f0ap3}).
For $A_t \ge A_t^{suf}$, this cubic equation in $H_2$ has
necessarily three real positive roots [see sec.3.3]. The
intermediate solution, denoted $H_2^{(0)}$, which can be given an
explicit analytical expression, always proves to be very close to
the CCB VEV $<H_2>$ (the discrepancy is less than $1 \%$).\\
 - Solve the extremal equation eq.(\ref{eqf3}) in $f$ with
$H_2=H_2^{(0)}$. This equation has only one consistent (i.e. real
and positive, see sec.2.2) solution $f^{(1)}$, which is even
closer to the CCB VEV $<f>$ than $f^{(0)}_3$.\\
 -The algorithm may be
iterated in the same way without ambiguity. The set of values
$(H_2^{(n)},f^{(n)})_{n\ge 0}$ proves to converge very fast
towards $(<H_2>,<f>)$.\\
 Once the CCB VEVs $<H_2>, <f>$ are computed , $<\tilde{t}_L>$ is
obtained by eq.(\ref{eqUL}) and we have $<\tilde{t}_R>= <f>
<\tilde{t}_L>$, which completes the determination of the location
of the CCB vacuum. The final step is the comparison of the
potential $V_3$, eq.(\ref{V3ch}), at this dangerous vacuum with
$<V>|_{EW}$. A scan for $A_t \ge A_t^{suf}$ then provides the
critical bound $A_{t,3}^c$.\\

For completeness, we have summarized the full algorithm  to
compute the critical bound $A_{t,3}^c$. It is however important to
stress that, in practice, this evaluation is considerably simpler
and more rapid. The values $(H_2^{(0)},f^{(0)}_3)$ obtained with
the first iteration of our algorithm already provide excellent
analytic approximations of $(<H_2>,<f>)$. Further iterations will
result in unimportant effects. In particular, the impact on the
critical CCB bound $A_{t,3}^c$ is extremely tiny, $\sim O(1 \
GeV)$. Thus, to an excellent accuracy, explicit analytic
expressions for all the VEVs of the CCB vacuum and of the
potential $V_3$, eq.(\ref{V3ch}), at this vacuum can be given, and
the determination of the critical CCB bound $A_{t,3}^c$
essentially reduces to the comparison of the CCB potential $V_3$,
eq.(\ref{V3ch}), at the CCB vacuum with the EW potential
$<V>|_{EW}$, eq.(\ref{VEW}).

\section{Numerical illustration of the CCB bounds}

We turn now to the numerical illustration of the various CCB
bounds obtained in the plane $(H_2, \tilde{t}_L, \tilde{t}_R)$,
first for low $\tan \beta$ [we take $\tan \beta=3$], and then for
large $\tan \beta$ [we consider the limiting case $\tan \beta=+
\infty$], where the additional negative contribution of the Higgs
mass parameter $m_2^2$ induces new features of the potential, as
shown in sec.3.4. In order to incorporate one-loop leading
corrections, the CCB bounds are implicitly supposed to be
evaluated at the SUSY scale $Q \sim Q_{SUSY}$. \\

{\bf $\bullet$ The low $\tan \beta$ regime}\\

In Figures 2-3, the behaviour of the CCB bounds on $A_t$, i.e. the
critical bound $A_{t,3}^c$, eq.(\ref{CCBcond}), the optimal
sufficient bound $A_t^{suf}$, eq.(\ref{condsuf}), which is always
equal $A_t^{(1)}$ in this regime, its approximation $A_t^{ap}$,
eq.(\ref{csufap}), and finally the traditional bound in the D-flat
direction $A_{t,3}^D$, eq.(\ref{condfrere}), is illustrated as a
function of the soft squark masses
$m_{\tilde{t}_L},m_{\tilde{t}_R}$. The set of values chosen is
consistent with a correct tree-level EW symmetry breaking with
$\tan \beta=3, m_{A^{0}}=520 \ GeV$ and a top quark mass $m_t=175
\ GeV$. As can be seen, in both illustrations, the hierarchy
$A_{t,3}^c \ge A_t^{suf}=A_{t}^{(1)}$, eq.(\ref{compsuf}), is
verified, as expected.\\
 In Figure 2, the various CCB bounds are
plotted as a function of the ratio
$r_1=m_{\tilde{t}_L}/m_{\tilde{t}_R}$, taking $m_{\tilde{t}_R}=200
\ GeV$. Therefore, except for $r_1=1$, the CCB vacuum always
deviates from the $SU(3)_c$ D-flat direction. Comparing the
critical bound $A_{t,3}^{c}$ and $A_t^{(1)}$, we see that they
follow each other closely for all values of $r_1$, with
$A_{t,3}^{c} \sim 1.04-1.10 \ A_t^{(1)} $, the lowest values being
reached for large $r_1$ and the largest for $r_1 \sim 0$. For $ 1
\lsim r_1 \le 5$, the sufficient bound $A_t^{(1)}$ is
approximately linear in $r_1$ and the accuracy of the
approximation $A_t^{ap}$ is rather good, better than $5 \%$.
Although this linear behaviour breaks down for low $r_1 \lsim 1$,
$A_t^{ap}$ still provides a good thumbrule to evaluate $A_t^{(1)}$
(within $5-8 \%$). Note that we can have either $A_t^{(1)} \ge
A_t^{ap}$ or $A_t^{(1)} \le A_t^{ap}$, showing that $A_t^{ap}$ is
just an approximation and should be handled with care.\\
\begin{figure}[htb]
\vspace*{-0.8cm}
\begin{center}
\mbox{ \psfig{figure=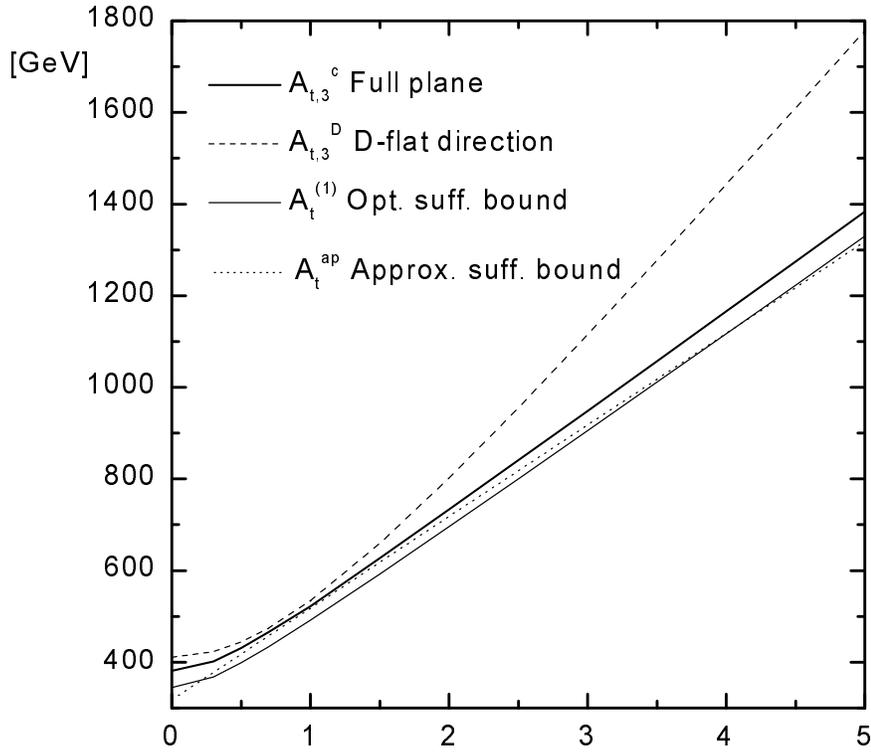,width=13cm}}
\end{center}
\vspace*{-1.8cm} \caption[Figure 2a]{{\sl CCB bounds versus
$r_1=m_{\tilde{t}_L}/m_{\tilde{t}_R}$. We take $m_{\tilde{t}_R}=
200 \ GeV, \ m_1= 400 \ GeV, \ m_3= \ 228 \ GeV$ and $m_2,Y_t,g_1,
g_2, g_3$ as in Fig.1 . This gives $m_t=175 \ GeV$.}}
\end{figure}
For $r_1 \sim 1$, we have $A_{t,3}^D \sim A_t^{(1)} \sim
A_{t,3}^c$. In this regime, the soft squark masses are of the same
order, implying that the CCB vacuum is nearly aligned in the
$SU(3)_c$ D-flat direction, as noted in sec.3.4. The CCB vacuum is
also located in the vicinity of the $SU(2)_L \times U(1)_Y$ D-flat
direction, because the common value of the soft squark masses
$m_{\tilde{t}_L} \sim m_{\tilde{t}_R} \sim 200 \ GeV$ is not so
large compared to the Higgs mass parameter $m_2=118 \ GeV$, so
that the relation eq.(\ref{massrel}) is approximately verified.
For large $r_1 \sim 5$ and $A_t \sim A_{t,3}^c$, the CCB vacuum is
located far away from the $SU(3)_c$ D-flat direction (as well as
the $SU(2)_L \times U(1)_Y$ D-flat direction). This large
departure clearly appears by comparing the traditional CCB bound
in the D-flat direction, $A_{t,3}^D \sim 1778 \ GeV$, and the
critical bound $A_{t,3}^c \sim 1383.5 \ GeV$. The latter is about
$30 \%$ below the traditional bound $A_{t,3}^D$! This is a typical
feature of this D-flat direction condition: for large soft squark
masses, it is far from being optimal and not very indicative of
the critical bound $A_{t,3}^c$. A better estimate is given by the
optimal sufficient bound $A_t^{(1)}$ or even its approximation
$A_t^{ap}$.\\
\begin{figure}[htb]
\vspace*{-0.8cm}
\begin{center}
\mbox{ \psfig{figure=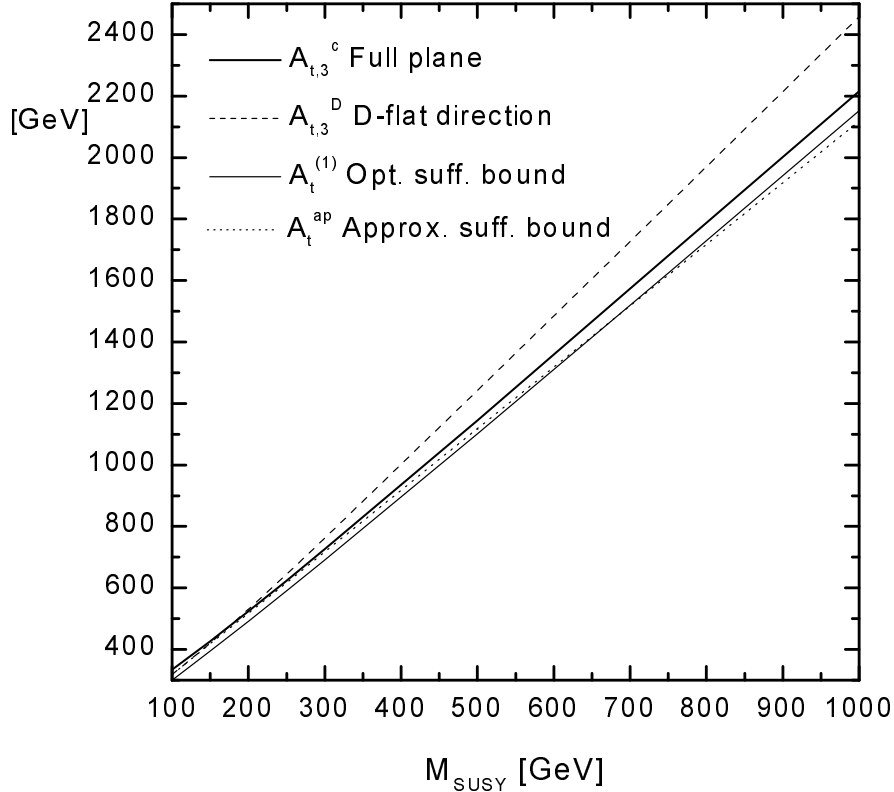,width=13cm}}
\end{center}
\vspace*{-1.6 cm} \caption[Figure 2b]{CCB bounds versus
$M_{SUSY}=m_{\tilde{t}_L}=m_{\tilde{t}_R}$. Same set of values as
in Fig.1-2 for the other parameters.}
\end{figure}
 Finally, we note that if we had taken
$r_{-1}=m_{\tilde{t}_R}/m_{\tilde{t}_L}$ and $m_{\tilde{t}_L}=200
\ GeV$, the curves obtained would overlap the ones presented here.
This is obviously an exact result for $A_{t,3}^D,A_t^{ap}$ [see
eqs.(\ref{condfrere},\ref{csufap})], but it proves also to occur
to a very good approximation for $A_{t,3}^c,A_t^{(1)}$.\\
 In Figure 3, the various CCB bounds are now plotted as a  function of
$M_{SUSY}=m_{\tilde{t}_L}=m_{\tilde{t}_R}$, with the same set of
values as in Fig.2 for the other parameters. The CCB vacuum is now
automatically aligned in the $SU(3)_c$ D-flat direction, but not
in the $SU(2)_L \times U(1)_Y$ D-flat one, except for
$M_{SUSY}=m_2=118 \ GeV$, see eq.(\ref{massrel}).\\
 In this illustration, we recover the same
qualitative behaviour of the CCB bounds as in Fig.2. Comparing
this illustration with the previous one for an equal value of
$M_{SUSY}^2=(m_{\tilde{t}_L}^2+m_{\tilde{t}_R}^2)/2$, we
furthermore observe that the difference $|A_{t,3}^D -A_{t,3}^c|$
is smaller in Fig.3 than in Fig.2, precisely because of this
alignment in the $SU(3)_c$ D-flat direction. In Fig.2, for
instance, for $M_{SUSY}^2=(720 \ GeV)^2$ with $r_1=5$, the
critical bound $A_{t,3}^c$ is about $ 22 \%$ below the traditional
bound $A_{t,3}^D$, whereas for an equal value of $M_{SUSY}=720 \
GeV$ in Fig.3, which gives the same value for $A_{t,3}^D$, the
critical bound $A_{t,3}^c$ is now just about $10 \%$ below
$A_{t,3}^D$. This illustrates the fact that any departure of the
CCB vacuum from the $SU(3)_c$ D-flat direction or, equivalently,
any splitting  between the soft squark masses, tends to lower
substantially the critical bound $A_{t,3}^c$ below which there is
no CCB danger.\\

 {\bf $\bullet$ The large $\tan \beta$ regime}\\

\begin{figure}[htb]
\vspace*{-0.8cm}
\begin{center}
\mbox{ \psfig{figure=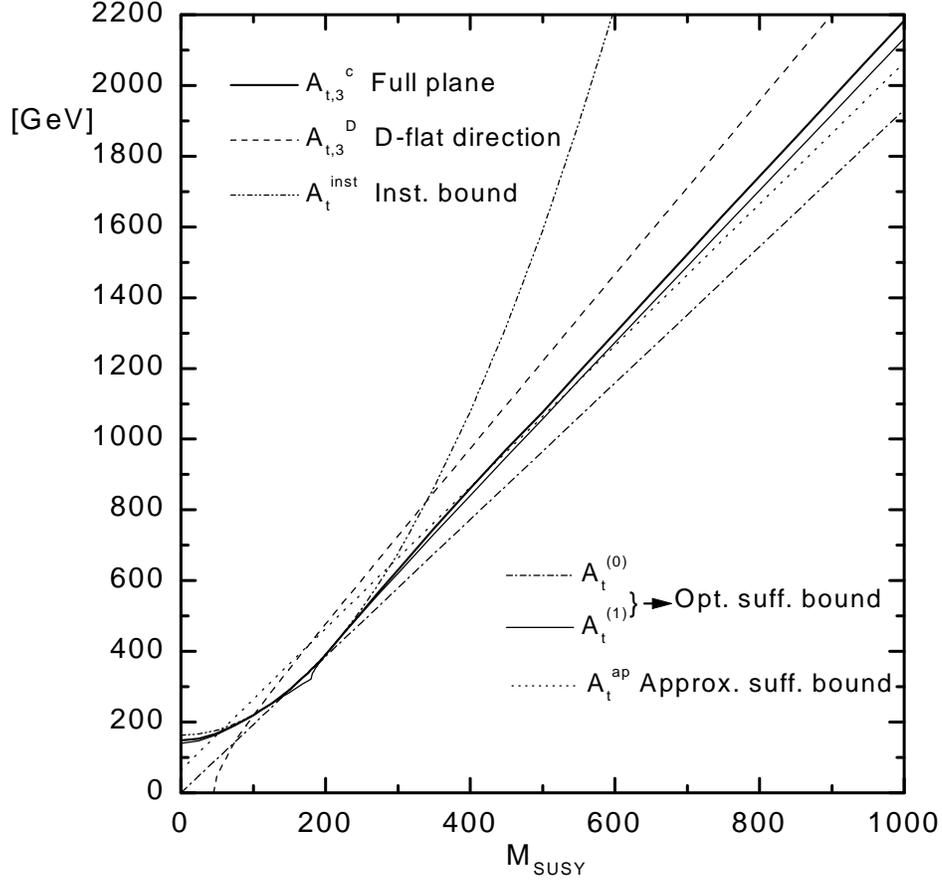,width=13cm}}
\end{center}
\vspace*{-2cm} \caption[Figure 3]{CCB bounds versus
$M_{SUSY}\equiv m_{\tilde{t}_L}=m_{\tilde{t}_R}$, for
 $\tan \beta=+ \infty$. The optimal sufficient bound is
$A_t^{suf} \equiv Max[A_t^{(0)},A_t^{(1)}]$. We take
 $m_2^2=-m_{Z^0}^2/2$ [see eq.(\ref{eqextre})] and  $m_t=175 \ GeV$
which implies $Y_t=1$.}
\end{figure}
Figure 4 is devoted to the large $\tan \beta$ regime. We take
$M_{SUSY}\equiv m_{\tilde{t}_L}=m_{\tilde{t}_R}$, with $\tan \beta
=+ \infty$. This benchmark scenario is often considered in Higgs
phenomenology \cite{higgs,higgsrev} and this illustration is
presented to set the stage for the next section where the impact
of the CCB conditions on the stop mass spectrum and on the
one-loop upper bound on the lightest Higgs boson  mass, $m_h$,
will be considered. As will be shown in a forthcoming article
\cite{CCB4champs}, this extreme $\tan \beta$ case also proves to
be numerically representative of the large $\tan \beta$ regime,
i.e. $\tan \beta \gsim 15$, with furthermore $|\mu| \lsim
Min[m_{A^0},M_{SUSY}]$.\\
 In this benchmark scenario, the CCB
vacuum is automatically aligned in the $SU(3)_c$ D-flat direction.
Obviously, any discrepancy between the soft mass terms
$m_{\tilde{t}_L},m_{\tilde{t}_R}$ would induce a deviation from
this direction and, on the other hand, a numerical modification of
the CCB bounds illustrated here, but the qualitative behaviour of
the CCB bounds would remain the same. \\
 For $\tan \beta =+\infty$, the EW vacuum is trapped in the plane
$(H_2, \tilde{t}_L,\tilde{t}_R)$ and the depth of the EW potential
is determined to be $<V>_{EW}=-m_{Z^0}^4/2(g_1^2+g_2^2)$ [see
sec.3.4]. Figure 4 illustrates how this geometrical feature of the
potential affects the various CCB bounds, including the sufficient
bound $A_t^{(0)}$, eq.(\ref{condsuf}), and the instability bound
$A_t^{inst}$, eqs.(\ref{Ameta1},\ref{Ameta2}).\\
 For all $M_{SUSY}$, the hierarchy
$A_t^{suf} \equiv Max[A_t^{(0)},A_t^{(1)}] \le A_{t,3}^c \le
A_t^{inst}$ is respected, even in the critical region
$M_{SUSY}\sim m_t$. This hierarchy merely reflects the fact that,
on one hand, no CCB vacuum may develop for $A_t \le A_t^{suf}$,
implying $A_{t,3}^c \ge A_t^{suf}$. On the other, if $A_t \ge
A_t^{inst}$, the EW vacuum would be automatically unstable and
would bend down in the direction of a deeper CCB vacuum, which
implies necessarily $A_t^{inst} \ge A_{t,3}^c,A_t^{suf}$.\\
 For all $M_{SUSY}$, the critical bound $A_{t,3}^c$ is just above the
optimal sufficient bound $A_t^{suf}$. For instance, for $M_{SUSY}
\ge 210 \ GeV$, we have $A_t^{suf} \le A_{t,3}^c \lsim 1.02 \
A_t^{suf}$, with $A_t^{suf}=A_t^{(1)}$. This shows once again how
an accurate approximation $A_t^{suf}$ can be for the critical CCB
bound $A_{t,3}^c$. We note also that $A_t^{ap}$ provides a good
estimate of $A_t^{suf}$, at least for $M_{SUSY} \ge 300 \ GeV$.\\
 The traditional bound $A_{t,3}^{D}$ in the D-flat direction
exists only for $M_{SUSY} \gsim 45.56 \ GeV$, and above this value
it increases fast. Let us remark that for $M_{SUSY} \lsim 100 \
GeV$, $A_{t,3}^{D}$ is not a necessary upper bound on $A_t$ to
avoid CCB, because for $A_t \in [A_{t,3}^D,A_{t,3}^c]$, the CCB
vacuum is not deeper than the EW vacuum and is therefore not
dangerous. The traditional bound $A_{t,3}^D$ provides a necessary
condition to avoid CCB only for $M_{SUSY} \gsim 100 \ GeV$, but in
this case it is far from being sufficient. Typically much larger
than the critical CCB bound $A_{t,3}^c$, it should also be handled
with care. For instance, for $M_{SUSY} \sim 300 \ GeV$, it allows
for some values of $A_t$ above the instability bound
$A_{t}^{inst}$, implying a tachyonic lightest stop mass!\\
 For $110 \ GeV \lsim M_{SUSY} \lsim 210 \ GeV$, the various CCB bounds
$A_t^{(0)}, A_t^{(1)}, A_{t,3}^c, A_t^{inst}$ cluster. The
potential enters in the critical regime where the CCB vacuum
interfere with the EW vacuum. In this regime, the critical CCB
bound $A_{t,3}^c$ coincides with the instability bound
$A_t^{inst}$, implying that a dangerous CCB vacuum may develop
only if the lightest physical stop gets tachyonic. Included in
this small region of the parameter space, more precisely for $140
\ GeV \lsim M_{SUSY} \lsim 192 \ GeV$, the typical hierarchy
$A_t^{(1)} \ge A_t^{(0)}$ is violated and we have therefore
$A_{t}^{suf}=A_t^{(0)}$. We note however that the maximal
discrepancy between $A_t^{(0)}$ and $A_t^{(1)}$ is quite small,
less than $5 \ GeV$.\\
 Finally, we observe that for $M_{SUSY} \gsim
300 \ GeV$, the critical CCB bound $A_{t,3}^c$ is much lower than
the instability bound $A_{t}^{inst}$. This result has an important
physical consequence in the limiting case $\tan \beta=+ \infty$
considered here, for which the stop mixing parameter $\tilde{A}_t$
coincides with the trilinear soft term $A_t$. It implies that the
CCB critical bound $A_{t,3}^c$ provides stringent restrictions on
the mass spectrum of the stop quark fields. This important point
is addressed in the next section.

\section{The stop CCB maximal mixing}

We investigate in this section some physical implications of the
critical CCB bound on the stop mass spectrum and the one-loop
upper bound on the lightest Higgs boson  mass. We consider the
benchmark scenario $M_{SUSY}=m^2_{\tilde{t}_L}=m^2_{\tilde{t}_R}$,
with $\tan \beta=+ \infty$ \cite{higgs,higgsrev}. As noted in the
last section, this extreme $\tan \beta$ regime is also quite
representative numerically of the large $\tan \beta$ regime with
small $\mu$, i.e. $\tan \beta \gsim 15$ and $|\mu| \lsim
Min[m_{A^0},M_{SUSY}]$ \cite{CCB4champs}. To be optimal, the
extension to the low $\tan \beta$ regime, valid for all values of
$\mu$, requires an investigation of the extended plane $(H_1, H_2,
\tilde{t}_L,\tilde{t}_R)$. This case will be presented in a
separate article \cite{CCB4champs}.\\
 In the benchmark scenario
considered here, the stop mixing parameter $\tilde{A}_t=A_t+\mu/
\tan \beta$ equals the trilinear soft term $A_t$ and the squared
stop mass matrix is given by the lower $2 \times 2$ matrix of
${\cal M}^2|_{\overline{EW}}$, eq.(\ref{matst1}), taking
$H_2=v_2$. Accordingly, the squared masses for the stop
eigenstates read \cite{MSSM}:
\begin{equation}
\label{mstop}
 m_{\tilde{t}_{1},\tilde{t}_2}^2=
M_{SUSY}^2+ m_t^2 -{\frac{1}{4}} m_{Z^0}^2
 \mp {\frac{1}{2}} \sqrt{4 m_t^2
A_t^2+{\frac{(8 \ m_{W^{\pm}}^2 - 5 \ m_{Z^0}^2)^2}{36}}}
\end{equation}
\begin{figure}[htb]
\vspace*{-0.8cm}
\begin{center}
\mbox{ \psfig{figure=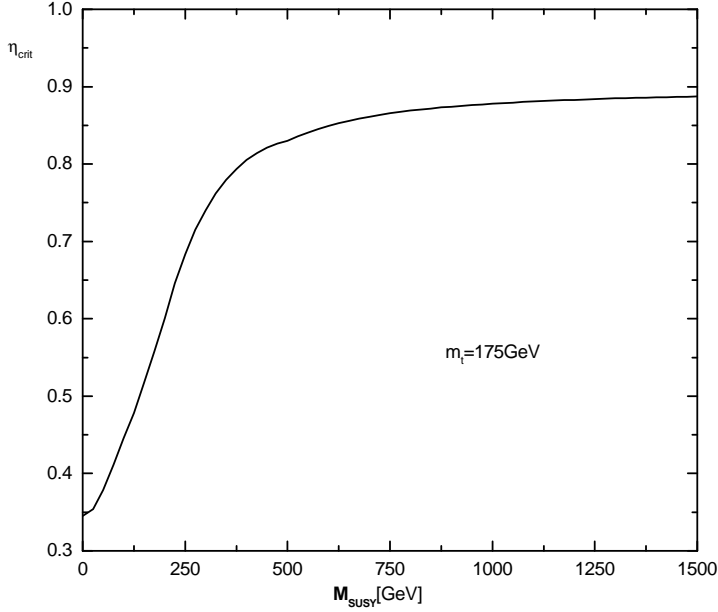,width=13cm}}
\end{center}
\vspace*{-1cm} \caption[Fig4a]{Critical CCB parameter $\eta_{crit}
\equiv A_{t,3}^c/A_t^{mix}$ versus $M_{SUSY}$ for $\tan \beta=+
\infty$. Same set of parameters as in Fig.4.}
\end{figure}
These masses depend only on two free parameters, $A_t$ and the
unified soft squark mass $M_{SUSY}$. However, taking into account
the CCB condition, a non-trivial correlation appears between these
two parameters. To avoid CCB in the plane $(H_2, \tilde{t}_L,
\tilde{t}_R)$, it is necessary and sufficient that $A_t \le
A_{t,3}^c$, eq.(\ref{CCBcond}), where the dependence in $M_{SUSY}$
of the critical bound $A_{t,3}^c$ is plotted in Fig.4. In the
following, all parameters are supposed to be evaluated at the
appropriate renormalization scale $Q \sim Q_{SUSY}$, in order to
incorporate one-loop leading corrections to this CCB bound.\\
 In Figure 5, we compare the critical bound $A_{t,3}^c$ to the
so-called Higgs maximal mixing $\tilde{A}_t^{max}$ commonly
considered in Higgs phenomenology \cite{higgs,higgsrev}:
\begin{equation}
\label{mixmax}
 \tilde{A}_t^{max} =A_t^{max} =\sqrt{6}
m_{\tilde{t}} \ \ with \ \ \ m_{\tilde{t}}^2=M_{SUSY}^2+m_t^2
\end{equation}
As is well-known, the lightest Higgs boson  mass, $m_h$, receives
a large one-loop correction arising from top and stop loops,
proportional to $m_t^4$ and which grows logarithmically with
$M_{SUSY}$. This correction is essential to overcome the
tree-level upper bound $m_h \le m_{Z^0}$ and is maximized for the
Higgs maximal mixing, eq.(\ref{mixmax}) \cite{higgs,higgsrev}. We
stress however that there is no physical reason to exclude a stop
mixing larger than this one, provided the masses obtained for the
lightest stop and CP-even Higgs boson  are not ruled out by
experimental data.\\
 The prominent fact in Fig.5 is that
$\eta_{crit}\equiv A_{t,3}^c/A_t^{max}$ is well below unity. We
have $0.35 \le \eta_{crit} \le {\frac{8}{9}} \sim 0.89$ for
$M_{SUSY} \le 1500 \ GeV$, showing that the CCB critical bound is
at least $10 \%$ below the Higgs maximal mixing,
eq.{\ref{mixmax}). {\sl Thus, the Higgs maximal mixing is always
ruled out by CCB considerations!} Moreover, it can be shown that
this striking result holds not only for large $\tan \beta$, but is
also typically verified for low $\tan \beta$ \cite{CCB4champs}.
Actually, the lower $\tan \beta$ is and the more CCB conditions
will tend to rule out such a large stop mixing.\\
\begin{figure}[htb]
\vspace*{-0.8cm}
\begin{center}
\mbox{ \psfig{figure=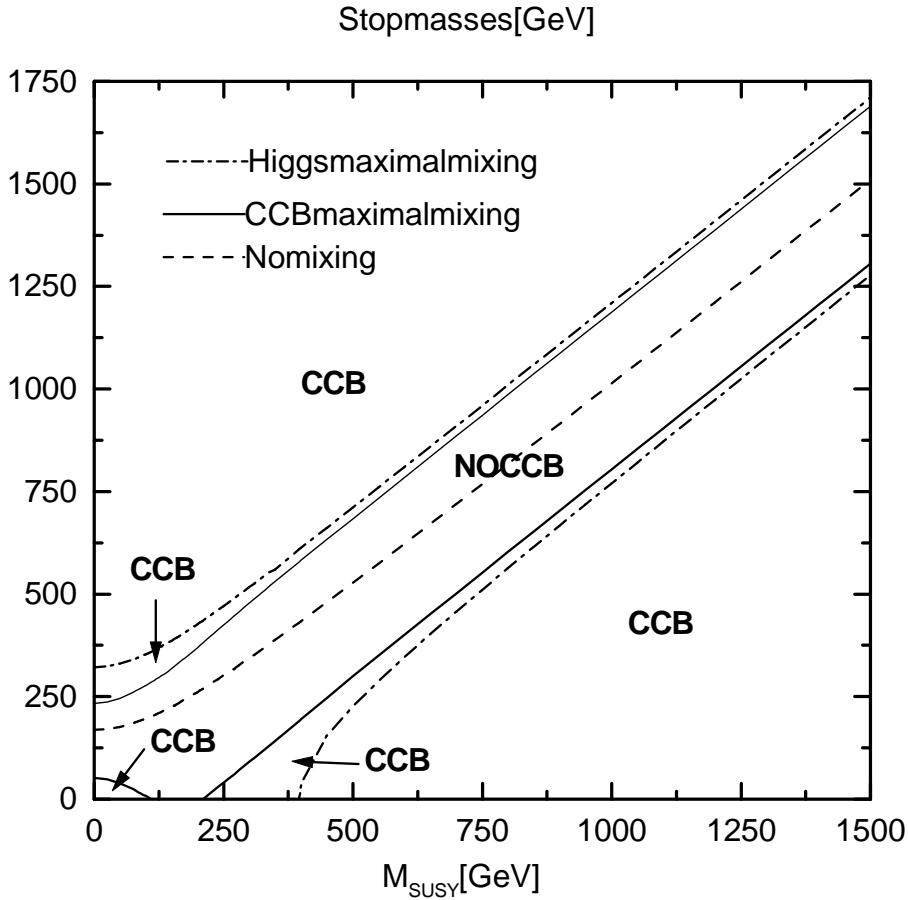,width=13cm}}
\end{center}
\vspace*{-1.8cm} \caption[Figure 4b]{Exclusion domain for the stop
masses versus $M_{SUSY}$, for $\tan \beta=+\infty$. The higher
curves [above the no mixing curve] provide upper bounds on the
mass of the heaviest stop $\tilde{t}_2$, and the lower [below the
no mixing curve] lower bounds on the mass of the lightest stop
$\tilde{t}_1$. Same set of parameters as in Fig.4.}
\end{figure}
 In the light of this new result, we introduce a new quantity, the
{\sl "CCB maximal mixing"}, defined to be the largest stop mixing
$\tilde{A_t}$ allowed by CCB considerations. Obviously, in the
case considered here, the CCB maximal mixing coincides with the
critical value $A_{t,3}^c$, plotted in Fig.4. Such a maximal
mixing has a clear physical meaning, which implies in particular
that the lightest Higgs boson  mass, $m_h$, cannot reach its
maximal value [at one-loop level, for the Higgs maximal mixing,
eq.(\ref{mixmax})], unless the EW vacuum is a metastable vacuum.\\
In Figure 6, we illustrate the bounds on the stop masses induced
by the CCB maximal mixing as a function of $M_{SUSY}$, and compare
them to the Higgs maximal mixing and also to the no mixing
($A_t=0$) cases. The lower curves correspond to the minimal values
allowed for the mass of the lightest stop $\tilde{t}_1$ and the
upper to the maximal values allowed for the mass of the heaviest
stop $\tilde{t}_2$.\\
 As expected, the bounds on the stop masses
induced by the CCB maximal mixing are more restrictive than for
the Higgs maximal mixing. All stop mass values compatible with a
Higgs maximal mixing are always located in the dangerous CCB
region. For $M_{SUSY} \lsim 425 \ GeV$, the Higgs maximal mixing
is already ruled out, either because it gives a tachyonic lightest
stop [for $M_{SUSY} \lsim 400 \ GeV$], or because the lightest
stop is too light [we take conservatively $m_{\tilde{t}_1} \gsim
100 \ GeV$] and should have been already found experimentally
\cite{stop1}. In this region of small $M_{SUSY}$, the CCB maximal
mixing enables us to avoid such a tachyonic lightest stop mass.
For $M_{SUSY} \lsim 110 \ GeV$, the lower bound on the lightest
stop mass slowly decreases with $M_{SUSY}$ and becomes exactly
zero for $110 \ GeV \lsim M_{SUSY} \lsim 210 \ GeV$. In this
critical region, the CCB vacuum interferes with the EW vacuum and
cannot be deeper than the latter, unless the lightest stop mass
becomes tachyonic.\\
 For $M_{SUSY} \lsim 310 \ GeV$, even the CCB maximal mixing is
excluded by the conservative experimental bound $m_{\tilde{t}_1}
\gsim 100 \ GeV$. {\sl Therefore, in this region of low
$M_{SUSY}$, the EW vacuum is necessarily the deepest one and
cannot be metastable.} This example illustrates how experimental
limits on the lightest stop combined with a precise determination
of the CCB condition may secure the EW vacuum in a large part of
the parameter space, so that metastability considerations become
completely irrelevant. In addition , we have also in this region
the upper bound $m_{\tilde{t}_2} \lsim 490 \ GeV$.\\
 For $M_{SUSY} \gsim  310 \ GeV$, the bounds on the stop spectrum
increase with $M_{SUSY}$. At $M_{SUSY}=500 \ GeV$, the
discrepancies between the CCB maximal mixing and the Higgs maximal
mixing cases are quite large for the lightest stop, $\Delta
m_{\tilde{t}_1} \sim 75 \ GeV$, and smaller but still important
for the heaviest stop $\Delta m_{\tilde{t}_2} \sim 30 \ GeV$. They
tend to decrease slowly with $M_{SUSY}$ and we have, e.g., $\Delta
m_{\tilde{t}_1} \sim 30 \ GeV$ and $\Delta m_{\tilde{t}_2} \sim 20
\ GeV$, for  $M_{SUSY}=1500 \ GeV$.\\ The linear behaviour of the
CCB bounds on the stop masses for large $M_{SUSY}$ is a direct
consequence of the asymptotic behaviour of the critical parameter
$\eta_{crit} \lsim 8/9 \sim 0.89$ [see Fig.5]. For $M_{SUSY} \gsim
500 \ GeV$, neglecting the gauge contributions, which are
unimportant in this regime, we obtain:
\begin{equation}
\sqrt{m_{\tilde{t}} ( m_{\tilde{t}} - {\sqrt{\frac{128}{27}}}
m_t)} \ \le m_{\tilde{t}_1} \le
 \ m_{\tilde{t}}\ \ \ \ \ , \ \ \ \ \ m_{\tilde{t}} \le m_{\tilde{t}_2}\  \le
 \
 \sqrt{m_{\tilde{t}} ( m_{\tilde{t}} + {\sqrt{\frac{128}{27}}} m_t)}
\end{equation}
giving indeed a linear behaviour for large $M_{SUSY}$:
$M_{SUSY}-\sqrt{{\frac{128}{108}}} m_t \le m_{\tilde{t}_1} \le
M_{SUSY}$ and $M_{SUSY} \le m_{\tilde{t}_2}\ \le \
M_{SUSY}+\sqrt{{\frac{128}{108}}} m_t$.\\

 Finally, we illustrate
the impact of the CCB maximal mixing on the CP-even lightest Higgs
boson mass $m_h$. At one-loop level, this mass has an upper bound
reached for $\tan \beta= + \infty$, $m_{A^0} \gg m_{Z^0}$, and the
Higgs maximal mixing, eq.(\ref{mixmax}), \cite{higgs,higgsrev}. We
have shown that the CCB condition rules out such a large stop
mixing, therefore this upper bound on $m_h$ may be lowered
\footnote{We note that the CCB maximal mixing, which coincides
with the critical bound $A_{t,3}^c$ in this case, does not depend
on the pseudo-scalar mass $m_{A^0}$. As will be shown in
\cite{CCB4champs}, this interesting property actually extends for
all values of $\tan \beta$ and $\mu$.}. For simplicity, we
consider this topic in a simplified setting, somewhat unrealistic,
taking only into account leading one-loop contributions coming
from top and stops loops. This will already point out the general
trend and the importance of CCB conditions in this context. In
this case, we have \cite{higgs,higgsrev}
\begin{equation}
\label{bmh} m_h^2 \le  m_{Z^0}^2 +{\frac{3 m_t^4}{4 \pi v^2}} [Log
{\frac{m_{\tilde{t}}^2}{m_t^2}}+ {\frac{A_t}{m_{\tilde{t}}^2}}(1-
{\frac{A_t}{12 m_{\tilde{t}}^2}})] \ \ \ , \ \ \ v=174 \ GeV
\end{equation}
\begin{figure}[htb]
\vspace*{-0.8cm}
\begin{center}
\mbox{ \psfig{figure=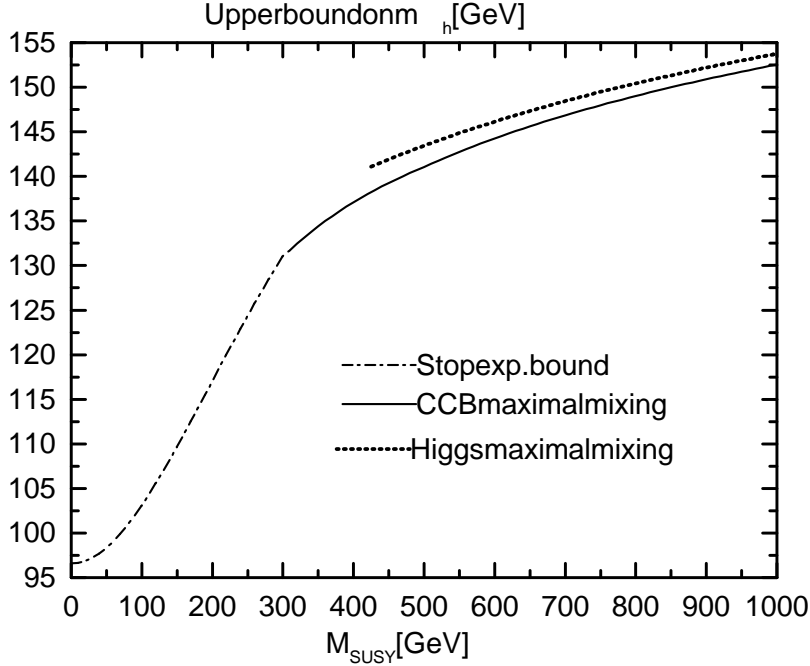,width=13cm}}
\end{center}
\vspace*{-1.8cm} \caption[Figure 4b]{Upper bound on the lightest
Higgs boson  mass versus $M_{SUSY}$, for $\tan \beta=+ \infty$. We
suppose $m_{\tilde{t}_{1}} \gsim 100 \ GeV$. Same set of
parameters as in Fig.4.}
\end{figure}
Up to one-loop level, it is consistent to combine the optimal
tree-level CCB condition $A_t \le A_{t,3}^c$, eq.(\ref{CCBcond}),
illustrated in Fig.4, with this upper bound. Figure 7 illustrates
the resulting one-loop CCB bound on $m_h$ as a function of
$M_{SUSY}$ and compares it with the maximal value reached for the
Higgs maximal mixing. As a direct consequence of the
aforementioned conservative experimental limit on the lightest
stop $m_{\tilde{t}_{1}} \gsim 100 \ GeV$ \cite{stop1}, three
regimes can be considered: i) For $0 \le M_{SUSY} \lsim 310 GeV$,
the lightest stop mass experimental bound is more stringent than
the CCB maximal mixing. In this regime, we have $m_h \le 131 \
GeV$. The experimental lower bound on $m_h$ puts in turn lower
bounds on $M_{SUSY}$. For instance, taking $m_h \ge 115 \ GeV$, we
must have $M_{SUSY} \gsim 190 \ GeV$. ii) For $310 \ GeV \lsim
M_{SUSY} \lsim 425 \ GeV$, the CCB maximal mixing becomes more
restrictive than the conservative lightest stop experimental mass
bound, while the Higgs maximal mixing is still irrelevant, either
because the lightest stop mass is tachyonic or too light. In this
regime, eq.(\ref{bmh}) combined with the critical CCB condition
$A_{t,3}^c$ [see Figs.4-5] give $131 \ GeV \le m_h \le 138 \ GeV$.
iii) For $M_{SUSY} \gsim 425 \ GeV$, the CCB and the Higgs maximal
mixing are both more restrictive than the experimental bound on
the lightest stop mass. The CCB upper bound on the lightest Higgs
boson mass is lower than the one for the Higgs maximal-mixing by
about $2.8 \ GeV$ for $M_{SUSY} \sim 425 \ GeV$, which is a rather
substantial effect. This discrepancy then decreases slowly for
larger $M_{SUSY}$. We have, e.g., respectively $\Delta  m_h \sim
(2.4,1.5,1.2) \ GeV$ for $M_{SUSY}=(500,750,1000) \ GeV$.\\
 A more realistic investigation of the consequences of CCB conditions on $m_h$
clearly requires that we go beyond the simple approximation given
by eq.(\ref{bmh}). The complete set of one-loop contributions
should be taken into acount, including in particular those arising
from bottom and sbottoms loops [which are themselves constrained
by strong CCB conditions for large $\tan \beta$,  see Appendix B]
\cite{higgsrev}. To leading order, we can however trust the
discrepancy on the upper bound of the lightest Higgs boson squared
mass between the Higgs and the CCB maximal mixing
\begin{equation}
\Delta m_h^2={\frac{9 (-1+\eta^2_{crit})^2 m_t^4}{4 \pi^2 v^2}}\ ,
\end{equation}
$\Delta m_h^2$ is actually independent of such additional
contributions and depends only of $M_{SUSY}$ via the critical
parameter $\eta_{crit}$.\\
 More importantly, two-loop contributions tend to lower
substantially this upper bound on the lightest Higgs boson  mass,
giving typically $m_h \lsim 130 \ GeV$ for $m_A, M_{SUSY} \gg
m_t$, with $\tan \beta \gg 1$ \cite{higgsrev}. Two-loop
non-logarithmic contributions are also responsible of a slight
displacement of the stop mixing value where this upper bound is
maximized \cite{higgsrev}. A refined study of the importance of
CCB conditions in this context, to be consistent at two-loop
level, should therefore require a complete one-loop level
investigation of CCB conditions in the plane
$(H_2,\tilde{t}_L,\tilde{t}_R)$, in order to take also into
account sub-leading effects induced by mass discrepancies in the
loops. Such a study is clearly beyond the scope of this article
and will the subject of further investigations. We believe however
that this simple illustration already clearly indicates the
crucial role CCB conditions can play in this phenomenological
context.

\section{Conclusions and outlook}

In this article, we have presented at the tree-level a complete
model-independent study of the CCB conditions in the plane
$(H_2,\tilde{t}_L, \tilde{t}_R)$. We have proposed a new procedure
to evaluate the CCB VEVs, which moreover enables us to obtain
excellent analytical approximations (at the level of the percent)
for the VEVs and, ultimately, for the optimal necessary and
sufficient conditions on $A_t$ to avoid CCB. The new conditions
incorporate the effect of all possible deviations of the CCB
vacuum from the D-flat directions, in particular from the
$SU(3)_c$ D-flat direction previously disregarded
\cite{CCB1,CCB2,CCB3,CCB4,CCB5}. We have pointed out that the CCB
vacuum typically deviates from the $SU(3)_c$ D-flat direction and
that this feature must be included in a consistent study of CCB
conditions to encompass the possibility of avoiding a tachyonic
lightest stop. This deviation is controlled essentially by the
discrepancy between the soft squark masses
$m_{\tilde{t}_L},m_{\tilde{t}_R}$. Rather small in an mSUGRA
scenario [where typically $m_{\tilde{t}_L} \sim m_{\tilde{t}_R}$],
it can be very large and make substantially more restrictive the
critical CCB conditions for $m_{\tilde{t}_L} \gg m_{\tilde{t}_R}$
or $m_{\tilde{t}_L} \ll m_{\tilde{t}_R}$. This should constrain
even more model-dependent scenarii, in particular those exhibiting
such large mass discrepancies at the SUSY scale, e.g. some anomaly
mediated models \cite{anom}, or, more generally, models
incorporating non-universalities for the squark soft masses of the
third generation at a high energy scale. In order to take into
account one-loop leading corrections, the tree-level CCB
conditions obtained in this article should be evaluated at an
appropriate scale $Q \sim Q_{SUSY}$, where $Q_{SUSY}$ is an
average of the SUSY masses.\\
 In the benchmark scenario $M_{SUSY}=m_{\tilde{t}_L}=m_{\tilde{t}_R}$ and
$\tan \beta = + \infty$, we have illustrated at this scale
$Q_{SUSY}$ some physical consequences of the critical CCB
condition in the plane $(H_2,\tilde{t}_L, \tilde{t}_R)$. A strong
bound on the stop mixing parameter $\tilde{A}_t$ [$=A_t$ in this
case] was obtained, ruling out by more than $10 \%$ the Higgs
maximal mixing $|A_t|= \sqrt{6} m_{\tilde{t}}$. This led us to
introduce a "CCB maximal mixing" for the stop fields. We have
exhibited new strong limits on the stop mass spectrum, which
simply encode the physical requirement of avoiding CCB. Finally,
we have considered the impact of the CCB maximal mixing on the
upper bound of the CP-even lightest Higgs boson mass, $m_h$, at
one-loop level, though in a simplified and rather unrealistic
setting. Taking into account only top and stop contributions, we
have shown that this upper bound can be reduced by up to $ \sim 3
\ GeV$ in comparison with the maximal value reached for the Higgs
maximal mixing. We believe that these illustrations stress the
importance of a refined study of CCB conditions, such as the one
presented here, in the context of Higgs phenomenology. We note
however that a more realistic investigation of the upper bound on
the lightest Higgs boson mass, $m_h$, requires that we take into
account all one-loop contributions to the Higgs mass, not only the
leading top and stop ones, but also two-loop contributions. As is
well-known, the latter can be large and are also responsible of a
displacement of the stop mixing which maximizes $m_h$
\cite{higgs,higgsrev}. A refined analysis of this important
phenomenological topic, to be consistent at two-loop level, should
therefore require a precise one-loop study of CCB conditions in
the plane $(H_2,\tilde{t}_L,\tilde{t}_R)$, in order to obtain
sub-leading contributions that our renormalization group improved
tree-level CCB conditions cannot grasp. Such a tedious sudy will
be the subject of future investigations.\\
 In the benchmark scenario considered in
this article, we have also pointed out that combining a precise
CCB information in the plane $(H_2, \tilde{t}_L,\tilde{t}_R)$ with
a conservative experimental imput on the lightest stop mass,
$m_{\tilde{t}_1} \gsim 100 \ GeV$, already indicates that the EW
vacuum is the deepest vacuum and is therefore stable in a large
part of the parameter space, $M_{SUSY} \le 310 \ GeV$. Similar
regions can also be found for any value of $\tan \beta $
\cite{CCB4champs}. Outside these regions, following the philosophy
of metastability, the EW vacuum can still be considered as safe,
even in the presence of a deeper CCB vacuum, provided its lifetime
exceeds the age of the Universe \cite{CCB5,CCB6}. A numerical
study of the tunneling rate into the CCB vacuum is required to
evaluate the relaxed CCB metastability condition \cite{meta}. The
present study, which can be straightforwardly completed by giving
accurate analytical expressions for the CCB saddle-point, provides
also some enlightening pieces of information on the shape of the
potential barrier between the vacua, and therefore give essential
tools to investigate precisely this feature.\\
 For completeness, it is important to stress that this investigation
of CCB condition in the plane $(H_2,\tilde{t}_L, \tilde{t}_R)$ is
also numerically illustrative of what can be found in the extended
plane $(H_1, H_2,\tilde{t}_L, \tilde{t}_R)$, provided $\tan \beta
\gsim 15$ and $|\mu| \lsim Min[m_{A^0},M_{SUSY}]$. In particular,
the results obtained for the critical CCB bound on $A_t$ and the
physical implications on the stop mixing parameter and the stop
mass spectrum are not substantially modified compared to the
extreme case $\tan \beta = + \infty$ illustrated here. In a
forthcoming paper, we will present the extension of this study to
the plane $(H_1, H_2,\tilde{t}_L, \tilde{t}_R)$, and give optimal
CCB constraints on $(A_t,\mu)$ valid for all values of $\tan
\beta$ \cite{CCB4champs}. We have also re-analyzed in a fully
model-independent way the potentially dangerous direction
$(H_1,H_2,\tilde{t}_L,\tilde{t}_R,\tilde{\nu}_L)$, previously
considered in \cite{CCB2}. Additional, though not very
restrictive, CCB conditions involving the sneutrino soft mass
$m_{\tilde{\nu}_L}$ will be given \cite{CCB5champs}.\\
 Besides physical implications on the MSSM mass
spectrum, the CCB condition on $A_t$, completed with the one on
the $\mu$-term obtained in the extended plane $(H_1, H_2,
\tilde{t}_L,\tilde{t}_R)$ \cite{CCB4champs}, should also have
further important consequences on the phenomenology of the MSSM
Higgs bosons \cite{pheno1}. In particular, CCB conditions provide
dramatic restrictions on physical processes which require, to be
competitive, a stop mixing parameter $\tilde{A_t}$ as large as the
Higgs maximal mixing, e.g., for the production of neutral Higgs
bosons associated with top squarks \cite{pheno2}. Further
investigations are currently made in this direction in order to
delineate more precisely the potential discovery of
Supersymmetry.\\

\noindent{\bf Acknowledgement}\\

\noindent This work was supported by a Marie Curie Fellowship,
under contract No HPMF-CT-1999-00363. I would like to thank
especially G.J. Gounaris and P.I. Porfyriadis for their help and
the Theory Group of the University of Thessaloniki for its
hospitality. I would like also to thank G. Moultaka for valuable
discussions and support during the completion of this work, as
well as for reading the manuscript and helping me to improve it.
This work was initiated in the French GDR in Supersymmetry. I
thank its participants for stimulating discussions. Special thanks
also to M. Bezouh.\\

\renewcommand{\theequation}{A.\arabic{equation}}
\renewcommand{\thesection}{A.\arabic{section}}
\setcounter{equation}{0} \setcounter{section}{0}

{\large \bf Appendix A: The optimal sufficient bound}\\

In this appendix, we give some details on the derivation of the
optimal sufficient bound $A_t^{suf}$ defined in sec.3.3, see
eq.(\ref{newsuf}). Let us consider the extremal equation
associated with $H_2$, eq.(\ref{eqH23}):
\begin{equation}
E_{H_2}= \alpha_3 H_2^3+ \beta_3 H_2^2+ \gamma_3 H_2+\delta_3=0
\nonumber
\end{equation}
where the coefficients $\alpha_3,\beta_3,\gamma_3,\delta_3$ are
given in the text, eqs.(\ref{alpha}-\ref{delta}). Obviously,
$\delta_3$ is always positive; $\beta_3$ is proportional to the
positive coefficient of the quadratic term in $<H_2>$ of $B_3$,
eq.(\ref{B3}), and is also positive; $\alpha_3$ is dominated by a
large contribution in $Y_t^4$ and is therefore negative; finally,
$\gamma_3$ is dominated by the negative terms proportional to
$A_t^2$ and $m_{\tilde{t}_L}^2+f^2 m_{\tilde{t}_R}^2$ and is
negative. The sign of these coefficients imply that any real root
of $E_{H_2}$, considered as a cubic polynomial in $H_2$, must be
positive. Moreover, by simple inspection of these roots expressed
as a function of the coefficients
$\alpha_3,\beta_3,\gamma_3,\delta_3$, it is straightforward to
show that the necessary and sufficient condition to have only one
real root is indeed given by eq.(\ref{condsuf1}).\\
 Let us prove now that the potential $V_3$,
eq.(\ref{V3ch}) has no local CCB minimum, if the extremal equation
associated with $H_2$, eq.(\ref{eqH23}), has only one real root in
$H_2$ for any value $f$.\\
 Equivalently, we can show that if the potential
$V_3$, eq.(\ref{V3ch}), has one local CCB minimum $(<H_2>$,
$<\tilde{t}_L>$,$<f>$), then for $f=<f>$ the extremal equation
$E_{H_2}=0$ has necessarily three real roots in $H_2$.\\
 Let us consider the following continuous path $\cal{P}$ in the plane
$(H_2,\tilde{t}_L,\tilde{t}_R)$: for $\phi  \in [0,\tilde{H}_2]$,
we take $H_2=\phi, \tilde{t}_L=\tilde{t}_R=0$; for $\phi \in
[\tilde{H}_2,<H_2>]$, we take $H_2=\phi$, $\tilde{t}_L=\sqrt{-2
B_3/A_3}|_{f=<f>}$, $\tilde{t}_R=<f> \tilde{t}_L$, where $A_3,
B_3$ are given in eq.(\ref{B3}). Here, the positive value
$\tilde{H}_2$ denotes the lowest solution of the equation $B_3=0$
for $f=<f>$. By definition, the path $\cal{P}$ goes through the
origin of the fields and also through the local CCB vacuum for
$\phi=<H_2>$ [see eq.(\ref{eqUL})]. \\
 We can show now that, in this
situation, it is absurd to have only one real root for the
extremal equation associated with $H_2$, eq.(\ref{eqH23}). By
definition, the solutions of this equation provide in particular
the value $H_2$ of any directional extremum in the second part of
the path $\cal{P}$, i.e. for $\phi \in [\tilde{H}_2,<H_2>]$.
Therefore, assuming that this equation has only one real root for
$f=<f>$ implies that on this part of the path $\cal{P}$, there is
no directional saddle-point. What about the first part of the path
$\cal{P}$, i.e. for $\phi  \in [0,\tilde{H}_2]$? Here, the
potential $V_3$, eq.(\ref{V3ch}), reads:
\begin{equation}
\label{V3a}
 V_3=H_2^2 (m_2^2+ {\frac{(g_1^2+g_2^2)}{8}}H_2^2)
\end{equation}
For $m_2^2 \ge 0$, this potential is monotonous as a function of
$H_2$ and has no non-trivial extrema. Hence, in this case, we
finally conclude that we can find a continuous path $\cal{P}$ in
the plane $(H_2,\tilde{t}_L,\tilde{t}_R)$ which connects the
origin of the fields and the CCB vacuum, moreover without any
directional saddle-point.\\
 We remind the reader that we
have assumed in our study positive squared soft mass
$m_{\tilde{t}_L}^2, m_{\tilde{t}_R}^2$ to avoid an obvious CCB
problem at the origin of the fields and that the CCB extremum
$(<H_2>$, $<\tilde{t}_L>$,$<f>$) we consider is supposed to be a
local minimum of the potential $V_3$, eq.(\ref{V3ch}). In this
light, the conclusion obtained for $m_2^2 \ge 0$ is absurd,
because in this case the origin of the fields is also a local
minimum, so that, necessarily there must be a barrier separating
it from the CCB vacuum,  and a saddle-point on any path connecting
them.\\
 The regime $m_2^2 \le 0$ is somewhat more complicated to investigate
and requires that we adjust the path $\cal{P}$ to different cases.
This comes from the fact that the potential in eq.({\ref{V3a}})
has an additional non-CCB extremum, $<H_2>_{\overline{EW}}^2 = -4
m_2^2/(g_1^2+g_2^2)$, as noted in sec.3.4. Therefore, the cases
$<H_2>  \ll <H_2>_{\overline{EW}}$ , $<H_2> \sim
<H_2>_{\overline{EW}}$, and $<H_2>  \gg <H_2>_{\overline{EW}}$
should be considered separately. The path $\cal{P}$ proposed is
obviously only adapted to the last case. We will not enter into
such a detailed, but straightforward, demonstration. Actually,
assuming that this additional non-CCB extremum is a local minimum,
as done in this article [see sec.3.4, eq.(\ref{Ameta1})], it is
easy to convince one-self that in all these cases a conclusion
similar to the one obtained for $m_2^2 \ge 0$ is obtained: it is
absurd to suppose that the extremal equation associated with
$H_2$, eq.(\ref{eqH23}), has only one solution in $H_2$ for
$f=<f>$, because on any path connecting this additional non-CCB
minimum and the CCB minimum, there should be a saddle-point which
necessarily would show as an additional real solution of this
equation.\\
 Hence, without loss of generality, we conclude that if
$E_{H_2}=0$ has only one real solution in $H_2$ for any value of
$f$, then the potential $V_3$, eq.(\ref{V3ch}), cannot have any
local CCB minimum. In such a situation, the unique solution of
$E_{H_2}=0$ found is spurious and located outside the compact
domain where $<\tilde{t}_L>^2 \ge 0$,
eqs.(\ref{compact1},\ref{compact2}).\\

\renewcommand{\theequation}{B.\arabic{equation}}
\renewcommand{\thesection}{B.\arabic{section}}
\setcounter{equation}{0} \setcounter{section}{0}

{\large \bf Appendix B: CCB conditions in the plane
$(H_1,\tilde{b}_L,\tilde{b}_R)$}\\

The procedure to evaluate the VEVs of the extrema of the potential
and the geometrical picture presented in this article hold also
for the tree-level potential in the plane $(H_1,
\tilde{b}_L,\tilde{b}_R)$, provided the bottom Yukawa coupling
$Y_b$ is large enough, or equivalently for large $\tan \beta$.
Here, $\tilde{b}_L$ and $\tilde{b}_R$ stand for the left and right
sbottom fields of the same generation, and $H_1$ is the neutral
component of the corresponding Higgs $SU(2)_L$ scalar doublet. In
this plane, the tree-level potential reads \cite{MSSM,CCB1}:
\begin{eqnarray}
\label{V3chb}
 \overline{V}_3&=&m_1^2 H_1^2+m_{\tilde{b}_L}^2 {\tilde{b}_L}^2+
m_{\tilde{b}_R}^2 {\tilde{b}_R}^2
 -2 Y_{b} A_b H_1 \tilde{b}_L \tilde{b}_R
+Y_b^2 (H_1^2 {\tilde{b}_L}^2+H_1^2 {\tilde{b}_R}^2+
{\tilde{b}_L}^2 {\tilde{b}_R}^2) \nonumber \\ &&
+{\frac{g_1^2}{8}} (H_1^2-{\frac{{\tilde{b}_L}^2}{3}}- {\frac{2
{\tilde{b}_R}^2}{3}})^2+{\frac{g_2^2}{8}}
(H_1^2-{\tilde{b}_L}^2)^2+ {\frac{g_3^2}{6}}
({\tilde{b}_L}^2-{\tilde{b}_R}^2)^2
\end{eqnarray}
This potential is similar to ${V}_3$, eq.(\ref{V3ch}). In fact,
redefining the fields and parameters of $\overline{V}_3$ as
follows \cite{MSSM,CCB1,CCB2}
\begin{eqnarray}
&& H_1 \rightarrow H_2 \ \ \ \ \ \ \ , \ \ \ \ \ \ \tilde{b}_{R/L}
\rightarrow \tilde{u}_{R/L} \nonumber \\
 && m_1 \rightarrow m_2 \ \ \ ,\ \ \
 m_{\tilde{b}_{R/L}}\rightarrow m_{\tilde{u}_{R/L}} \ \ \ , \ \ \
 Y_b \rightarrow Y_u
\end{eqnarray}
we recover $ V_3$, eq.(\ref{V3ch}), except for a minor difference
coming from the $U(1)_Y$ D-term.\\
 The relevant expressions to study CCB conditions in this plane are the
following ones. The VEV $<\tilde{b}_L>$ verifies:
\begin{equation}
\label{eqbL} \bar{A}_3 \ <\tilde{b}_L>^2+ 2 \bar{B}_3=0
\end{equation}
where
\begin{eqnarray}
 \bar{A}_3&\equiv& g_1^2(2 <f_b>^2+1)^2/18+g_2^2/2+2 g_3^2
(<f_b>^2-1)^2/3+4 Y_b^2 <f_b>^2 \\
 \bar{B}_3 &\equiv &<H_1>^2 \frac{[12 Y_b^2 (<f_b>^2+1) -3
g_2^2-g_1^2 (2 <f_b>^2+1) ]}{12} \nonumber
\\ &&
 -2 A_b Y_b <f_b> <H_1>+
m_{\tilde{b}_L}^2+<f_b>^2 m_{\tilde{b}_R}^2
\end{eqnarray}
with $f_b \equiv \tilde{b}_L/\tilde{b}_R$. Quite similarly to what
happens in the plane $(H_2, \tilde{t}_L,\tilde{t}_R)$, $\bar{B}_3$
tells us when the large bottom Yukawa coupling regime opens. This
occurs, whatever $<f_b>$ is, for $Y_b \ge Max[\sqrt{( g_1^2+3
g_2^2)/12},g_1/\sqrt{6}] \sim 0.3$ (at the EW scale). Using the
tree-level relation $m_t/m_b =(Yt /Y_b) \tan \beta \sim 35$
\cite{MSSM}, with $Y_t \sim 1$, this requires $\tan \beta \gsim
10.5$.\\
 In this large $\tan \beta$ regime, we obtain a sufficient
bound to avoid CCB in the plane $(H_1, \tilde{b}_L, \tilde{b}_R)$,
similar to $A_t^{(0)}$, eq.(\ref{condsuf}). It reads
\begin{equation}
\label{condsufb} A_b \le A_b^{(0)} \equiv m_{\tilde{b}_L}
\sqrt{1-{\frac{g_1^2}{6 Y_b^2}}}+m_{\tilde{b}_R}
\sqrt{{1-{\frac{(3 g_2^2+g_1^2)}{12 Y_b^2}}}}
\end{equation}
We note however that $A_b^{(0)}$ can be quite small for $Y_b \sim
\sqrt{( g_1^2+3 g_2^2)/12}$. \\
 The extremal equation associated with $<H_1>$ reads
\begin{equation}
\label{eqH23b} \bar{\alpha}_3 H_1^3+ \bar{\beta}_3
H_1^2+\bar{\gamma}_3 H_1+\bar{\delta}_3=0
\end{equation}
with
\begin{eqnarray}
\label{alphab}  \bar{\alpha}_3 &=&-36 Y_b^4 (f_b^2+1)^2+[3 g_3^2
(g_1^2+g_2^2)+ g_1^2 g_2^2] (f_b^2-1)^2 \nonumber \\ &&+ 6 Y_b^2
g_1^2 (2 f_b^4+6 f_b^2+1)+18 Y_b^2 g_2^2 (2 f_b^2+1)  \\
 \label{betab}
 \bar{\beta}_3&=&9 A_b Y_b f_b [12 (f_b^2+1) Y_b^2-(2 f_b^2+1)g_1^2-3 g_2^2 ]
 \\
 \label{gammab}
  \bar{\gamma}_3&=&-72 A_b^2 f_b^2 Y_b^2-3
(m_{\tilde{b}_L}^2+f_b^2 m_{\tilde{b}_R}^2) [12 Y_b^2 (f_b^2+1)-
g_1^2 (2 f_b^2+1)-3 g_2^2] \nonumber \\ && + m_1^2 [72 Y_b^2
f_b^2+g_1^2 (2 f_b^2+1)^2+9 g_2^2+12 g_3^2 (f_b^2-1)^2] \\
 \label{deltab}
\bar{\delta}_3 &=&36 A_b Y_b f_b (m_{\tilde{b}_L}^2+f_b^2
m_{\tilde{b}_R}^2)
\end{eqnarray}
The extremal equation associated with $f_b$ reads
\begin{equation}
\label{eqf3b} \bar{a}_3 f_b H_1^2+ \bar{b}_3 H_1+\bar{c}_3 f_b=0
\end{equation}
with
\begin{eqnarray}
\label{a3b}   \bar{a}_3&=&2 Y_b^2 [ (18 Y_b^2-12 g_3^2) (f_b^2-1)
+9 g_2^2- g_1^2 (4 f_b^2-1)] \nonumber \\
 &&+(f_b^2-1) [ g_1^2 g_2^2+3
(g_1^2+g_2^2) g_3^2]  \\
 \label{b3b}
 \bar{b}_3&=& A_b Y_b [12 g_3^2 (f_b^4-1)-9 g_2^2+
g_1^2 (4 f_b^4-1)]  \\
 \label{c3b}
  \bar{c}_3&=& - m_{\tilde{b}_L}^2 [ 36
Y_b^2+12 g_3^2 (f_b^2-1) +2 g_1^2 (2 f_b^2+1)] \nonumber \\ &&
+m_{\tilde{b}_R}^2 [ 36 f_b^2 Y_b^2-12 g_3^2 (f_b^2-1)+9 g_2^2+
g_1^2 (2 f_b^2+1)]
\end{eqnarray}
The VEVs $(<H_1>, <f_b>)$ of a consistent CCB vacuum have to
verify this set of coupled equations and must furthermore be
included in the compact domain where $<\tilde{b}_L>^2 \ \ge 0$.\\
 The recursive algorithm to compute the VEVs of the CCB
vacuum is identical to the one presented in the text for the top
Yukawa coupling regime. A convenient initial value to accelerate
the procedure is $f^{(0)}_{b,3}=\sqrt{{\frac{A_b^2+2
m_{\tilde{b}_L}^2-m_{\tilde{b}_R}^2}{A_b^2+2
m_{\tilde{b}_R}^2-m_{\tilde{b}_L}^2}}}$.\\ The optimal sufficient
bound $A_b^{suf}$ below which no CCB vacuum may develop in the
plane $(H_1, \tilde{b}_L, \tilde{b}_R)$ is always given by the
largest solution in $A_b$ of the equation
$\bar{{\cal{{C}}}}_3\equiv [2 \bar{\beta}_3^3-9 \bar{\alpha}_3
\bar{\beta}_3 \bar{\gamma}_3+ 27 \bar{\alpha}_3^2
\bar{\delta}_3]^2+ 4 [-\bar{\beta}_3^2+3 \bar{\alpha}_3
\bar{\gamma}_3]^3=0$, taking $f=f^{(0)}_{b,3}$.\\
 The potential $\overline{V}_3$, eq.(\ref{V3chb}),
can have at most three extrema: the origin of the fields, a CCB
local minimum and a CCB saddle-point. In particular, there is no
additional non-CCB extremum. Such a possibility would appear for
$m_1^2 \le 0$, which requires $\tan \beta \sim 0$. This is
obviously outside the large bottom Yukawa coupling regime and is,
moreover, ruled out by experimental data. Finally, the necessary
and sufficient bound $A_{b,3}^c$ is obtained by scanning the
region $A_b \ge A_b^{suf}$ and comparing the potential
$\overline{V}_3$ with the EW potential $<V>|_{EW}$,
eq.(\ref{VEW}). For large $\tan \beta$, which implies $Y_b \sim
1$, the appropriate renormalization scale to evaluate the
tree-level CCB conditions $A_b^{suf},A_{b,3}^c$ in order to
incorporate one-loop leading corrections, is the SUSY scale
$Q_{SUSY}$, quite similarly to the case $(H_2, \tilde{t}_L,
\tilde{t}_R)$.

\end{document}